**Modeling hydrogen integration in energy system models: Best practices for policy insights**

Muhammad Maladoh Bah[ab1], Sheng Wang[d], Mohsen Kia[b], Andrew Keane[cd], Terence O'Donnell[c,d]

[a]Sustainable Energy Authority of Ireland (SEAI), Hatch Street Upper, Dublin 2, Ireland
[b]UCD School of Electrical and Electronic Engineering, UCD, Belfield, Dublin 4, Ireland
[c]UCD Energy Institute, University College Dublin (UCD), Belfield, Dublin 4, Ireland
[d]School of Engineering, Newcastle University, Newcastle upon Tyne, United Kingdom

February 2025

**Abstract**

The rapid emergence of hydrogen in long-term energy strategies requires a broad understanding on how hydrogen is currently modelled in national energy system models. This study provides a review on hydrogen representation within selected energy system models that are tailored towards providing policy insights. The paper adopts a multi-layered review approach and selects eleven notable models for the review. The review covers hydrogen production, storage, transportation, trade, demand, modeling strategies, and hydrogen policies. The review suggests existing models would often opt for a simplified representation that can capture each stage of the hydrogen supply chain. This approach allows models to strike a balance between accuracy and preserving computational resources. The paper provides several suggestions for modeling hydrogen in national energy system models.

**Keywords:** energy system models, hydrogen supply chain

---

[1] Correspondence: muhammad.maladohbah@seai.ie

The authors thank Dr. Emer Dennerhy, Dr. Pádraig Daly, and Denis Neary from the Sustainable Energy Authority of Ireland (SEAI) for fruitful discussions and insights. This paper is an output of a joint SEAI and UCD hydrogen fellowship titled '*Development of Green Hydrogen in Ireland*' (M.B). Any remaining errors are the authors responsibility.

| Abbreviation | Meaning |
| --- | --- |
| AEL | Alkaline Electrolyser |
| ASU | Air Separation Unit |
| ATR | Auto Thermal Reformation |
| CCS | Carbon Capture and Storage |
| CCUS | Carbon Capture Utilisation, and Storage |
| CHP | Combined Heat and Power |
| EFMS | Energy Futures Modeling System |
| EMDE | Emerging Market and Developing Economies |
| GEC | Global Energy and Climate Model |
| HAR1 | Hydrogen Allocation Round 1 |
| HDSAM | Hydrogen Delivery Scenario Analysis Model |
| HMM | Hydrogen Market Module |
| HPD | Hydrogen Process Design |
| HSC | Hydrogen Supply Chain |
| HSCD | Hydrogen Supply Chain Design |
| IEA | International Energy Agency |
| IRA | Inflation Reduction Act |
| IRENA | International Renewable Energy Agency |
| LCOH | Liquid Organic Hydrogen Carrier |
| LTES | Long Term Energy Scenarios |
| MCH | Methylcyclohexane |
| NEMS | National Energy Modeling System |
| PEM | Proton Exchange Membrane |
| SMR | Steam Methane Reforming |
| SOE | Solid Oxide Electrolyser |

# 1. Introduction

Global concerns over the impact of rising temperatures has injected an urgency to accelerate the transition to a cleaner energy system. To reach a net-zero target by 2050 would necessitate a substantial increase in the deployment of mature renewable technologies and new emerging technologies (IEA, 2021a). In this context, hydrogen is gradually emerging as a potent energy carrier of a future decarbonised energy system (Maryam, 2017). Between 2017 and 2023, approximately 41 countries published hydrogen strategies or roadmaps detailing their plans for a domestic hydrogen sector (DECHEMA and acatech, 2024) (see Section 2.1).

Fundamentally, hydrogen is a non-toxic, colourless, odourless, and highly combustible gas. It is a versatile energy carrier that can be produced from all primary energy sources (i.e., water, solar, biomass, coal, gas, nuclear), stored and transported in various ways (Blanco et al., 2022). These characteristics position hydrogen as an ideal energy carrier to decarbonize hard-to-abate sectors. In 2022, global hydrogen use reached 95 Mt. The bulk of the global hydrogen demand was produced with natural gas without carbon capture, utilisation, and storage (CCUS) (62%), followed by coal (21%), by-product hydrogen (16%) and then low emission-hydrogen (0.7%) (IEA, 2023a). Hydrogen is widely utilized in refining, chemical industry, steel industry, and for special industrial applications. Despite hydrogen's notable potential, it remains a nascent energy technology and global uptake is hampered by several factors such as inadequate hydrogen infrastructure, high costs, demand uncertainty, and underdeveloped technical and regulatory standards (Maryam, 2017; Hanley et al., 2018; Li et al., 2019; IRENA, 2021).

The hydrogen system is complex due to the multitude of options at each stage of the supply chain. At the generation stage, hydrogen can be generated using various primary energy sources as feedstock and stored in either gaseous or liquid form. At the transportation stage, hydrogen can be transported using dedicated pipelines, tube trailer, ships, or railway. The hydrogen supply chain components are therefore interlinked and integrated within a broader energy system (Bolat and Thiel, 2014). It is therefore vital to develop robust tools and models to adequately capture the hydrogen system.

The emergence of hydrogen in long-term energy policy targets necessitates the need to understand the interaction of hydrogen in national energy systems and future growth trajectories. In a recent guideline report, IRENA highlights the intricate link between energy models, long-term energy scenarios (LTES), and national hydrogen strategies (IRENA, 2024a). The report asserts that in the development of hydrogen policies, energy models form the foundation. The results of energy models feed into LTES, which is then combined with stakeholder consultation to create strategic documents and ultimately guides policies. National energy policy planners may adopt a suit of in-house modeling tools or outsource modeling tasks to specialised consultancy firms. In the former, energy planners may opt to build-upon existing hydrogen structures in their national energy models, re-develop, or in some cases develop a hydrogen representation models and tools from scratch.

The hydrogen literature has grown over the past years; however, studies that encapsulate best practices on modeling hydrogen in national energy system models are limited. To date, the literature has predominantly revolved around a review of models that addresses the hydrogen supply chain (HSC) (Dagdougui, 2012; Bolat and Thiel, 2014; Maryam, 2017; Blanco et al., 2022; Riera et al., 2023). Hanley et al. (2018) for example, reviewed the role of hydrogen in future low-carbon pathways within global, multi-regional, and national energy system models. The focus of the paper was primarily oriented to understand the policy drivers and scenarios that sparked the emergence of hydrogen within energy modelling frameworks and its interaction with other low-carbon technologies. Among the key catalysts for the emergence of hydrogen in low-carbon pathways are decarbonisation policy targets, high levels of renewable electricity penetration, the role for hydrogen in hard to decarbonise sectors, among others.

Two recent papers have attempted to capture the best practices for modeling hydrogen in energy systems models. In a comprehensive study, (Zhang et al., 2025) reviewed a total of 125 papers to investigate the system integration of hydrogen within energy system models. The study identified gaps in modeling approaches and future research avenues. The key distinction between (Zhang et al., 2025) and this paper is the review methodology. In this study, we identify and directly scrutinize selected energy system models that have contributed directly or indirectly to national energy policies as opposed to reviewing empirical studies. Additionally, our review is oriented towards policy assessments, and we therefore determine if and how existing energy system models and tools account for hydrogen policies and provide a range of suggestions for policy modeling. In another study, Langer et al. (2024) compiled best practices from a workshop that was structured on three modeling frameworks (*PyPSA*, *AnyMOD*, and *Balmorel*). However, the paper is mainly confined to network modeling, electrolysers, and flexibility. Beyond hydrogen, modeling best practice exists for general energy system modeling (DeCarolis et al., 2017), alternative technologies such as energy storage (Bistline et al., 2020), and integrating variable renewable energy (VRE) into the energy system (Collins et al., 2017).

This paper seeks to identify global best practices for incorporating hydrogen in national energy system models. Our study is aimed at providing a holistic guideline for energy analysts on what aspects to consider when modeling the hydrogen sector and for researchers on potential gaps in modeling approaches. To address the objective, this study conducts a comprehensive review of national energy system models with hydrogen components. The outcome of the review is a set of suggestions that can be used for future modeling of hydrogen for policy relevant applications. We also recognize that the modeling stage will require trade-offs between policy direction, level of detail, and considering the limits of technical and human resources.

The remainder of the paper is structured as follows. Section 2 provides a brief background on the global status of hydrogen. Section 3 outlines the systematic review methodology. Section 4 presents and discusses key hydrogen modeling elements. Section 5 provides concise hydrogen modeling suggestions. Finally, section 6 concludes the study.

2. **Global hydrogen development**

**2.1 Global context**

Globally, hydrogen is gradually emerging as a potent energy carrier capable of decarbonising hard to abate sectors. The IRENA *World Energy Transitions Outlook* 2023 report highlights that under the 1.5°C scenario, hydrogen and its associated derivative products will abate 12% of CO2 emissions and constitute 14% of direct final energy consumption (IRENA, 2023). Presently, global installed hydrogen capacity stands at approximately 1.1 GW (as of 2023) and consumption reached 95 Mt (Hydrogen Council and McKinsey & Company, 2023; IEA, 2023a). Approximately 41 countries have published hydrogen strategies or roadmaps detailing their plans for a domestic hydrogen sector and long-term electrolysis capacity targets as Fig. 1 illustrates. In Europe, 17 countries have announced plans to install approximately 52 GW of electrolysis capacity. Broadly, hydrogen remains confined to traditional applications such as refining and as a feedstock to produce chemical products. The deployment of hydrogen in new applications such as transportation, industry, production of hydrogen fuels, and electricity remains stagnant (IEA, 2023a).

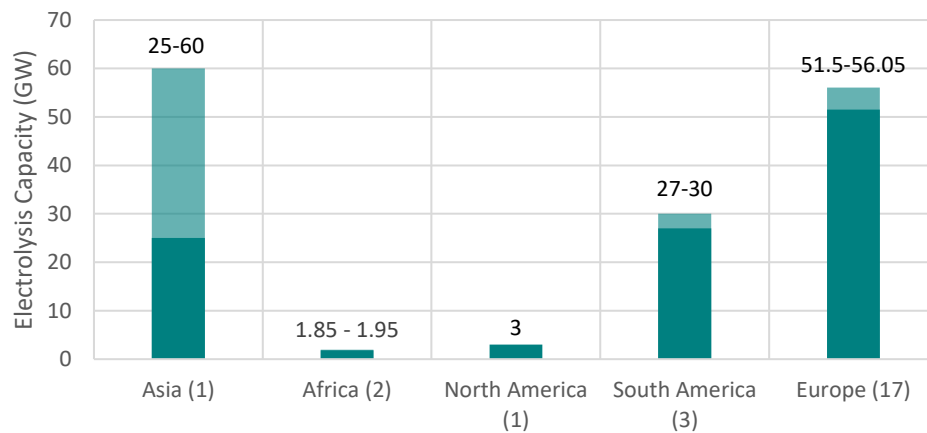

**Fig. 1.** Global electrolysis capacity by region in 2030
**Note:** Values in parentheses represents the number of reviewed countries in each region.
Source: (DECHEMA and acatech, 2024).

Currently, hydrogen production costs remain high, but are expected to decline over the coming decades. Frieden and Leker (2024) estimated that the cost of hydrogen production from electrolysis are expected to decline from a current baseline of €5.3/kg in 2020 to €4.4/kg in 2030 and €2.7/kg in 2050. Hydrogen production cost from unabated natural gas is the cheapest compared to renewable based sources at approximately €1.38 to €5.52/kg as Fig. 2 illustrates.

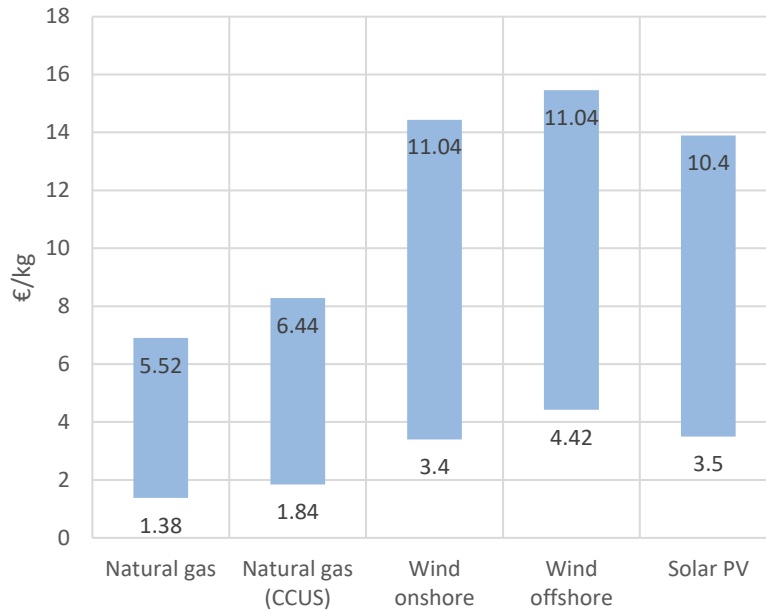

**Fig. 2**. Hydrogen production costs by energy source as of 2023.
Source: (IEA, 2023a; Frieden and Leker, 2024).

To bridge the cost gap between renewable hydrogen and fossil-fuel based hydrogen, several governments have introduced support mechanisms to stimulate the uptake of hydrogen. The U.S. federal government introduced two direct schemes for clean hydrogen producers in the 2022 Inflation Reduction Act (IRA). Namely, the clean hydrogen production tax credit (PTC) (26 U.S.C § 45V) and the tax credits for carbon oxide sequestration (26 U.S.C § 45Q). The UK introduced the hydrogen production business model in the Energy Act 2023. In the first hydrogen allocation round (HAR1), the Department for Energy Security and Net Zero (DESNZ) selected 11 out of 17 projects with a combined capacity of 125 MW (DESNZ, 2023). The HAR1 round will provide over £2 billion in revenue support to the selected projects. Similarly, the European Union (EU) launched the first pilot of the EU Hydrogen bank in November 2023. In the first auction round, seven projects were selected from a total of 132 projects. These projects will receive pay-as-bid revenue support over a 10-year period.

3. Review approach

To address the primary research objective of the study, this paper adopts a detailed two-layered approach. In the first layer, all countries that published national hydrogen strategies or have hydrogen plans were identified. The initial list of countries selected was based the (DECHEMA and acatech, 2024) annual report on international hydrogen strategies. The report identified 43 countries and regions with published hydrogen strategies as of January 2024. The list was cross-checked with reports from IRENA (IRENA, 2024b), World Energy Council and the International Hydrogen Progress Index (Hydrogen UK and ENA, 2023). Subsequently, the national hydrogen strategy for each country was reviewed to identify the modelling framework behind the assessments. In cases where the national hydrogen strategies did not make any reference to a modelling framework, the countries energy strategy documents and government reports were reviewed. Table 1 summarizes the national energy models that were captured in the first stage of the review.[2]

The second layer widens the net of potential energy system models and reviews hydrogen survey papers. This layer primarily draws on the findings of six papers on modelling hydrogen as highlighted in Table 2. Half of the reviewed papers were on modelling hydrogen supply chain using energy system models. For example, (Bolat and Thiel, 2014) conducted a comprehensive two-part review on bottom-up energy models of hydrogen systems. The focus of the first part of the study is on classifying 19 different hydrogen supply pathways that can be embedded into energy system models. (Li et al., 2019) conducted a broad review and classification of models for the hydrogen supply chain network. The author grouped models into three categories, namely, energy system optimization models, geographically explicit optimization models, and refuelling station locating models.

Riera et al. (2023) conducted a literature survey on the optimization models that are tailored to the hydrogen supply chain design (HSCD) and hydrogen process design (HPD). The authors reviewed 99 papers and classified the models into five broad categories (model type, uncertainty, spatial scale, investment stage, and objective). In a comprehensive paper, (Blanco et al., 2022) provided a tailored taxonomy for classifying hydrogen model archetypes. The authors surveyed 49 articles and 29 were incorporated in hydrogen taxonomy. The study classified hydrogen models into nine archetypes, each with its own set of models. Once suitable models were identified, the next step is to extract key hydrogen modeling features.

---

[2] From the initial list of 43 countries, we excluded countries that fail to disclose a modeling framework in the strategy documents or where the language barrier made it challenging to assess.

**Table 1:** Reviewed energy system models

| Model family /custom model | Country/organization |
|---|---|
| TIMES | Norway (Danebergs et al., 2022), Portugal (Gouveia et al., 2012), Ireland (Balyk et al., 2022), Scotland (Scottish Government, 2018; Dodds, 2021), Sweden (Riekkola, 2015), South Africa (Hughes et al., 2021), Russia (Lugovoy, 2007), Spain (Cabal et al., 2012), China, Switzerland (Kannan and Turton, 2014), Belgium (Laguna et al., 2022; VITO, 2013), India (SARI/EI, 2016), Poland, Italy (Cosmi et al., 2009; Gaeta, 2013), Germany, UK (Daly and Fais, 2014), Finland, Denmark (Balyk et al., 2019) |
| MARKAL | China |
| PyPSA | Poland (Czyżak et al., 2021) |
| Balmorel | Poland |
| PRIMES Model | EU Commission, Poland |
| EnergyPLAN | Spain |
| Canada Energy Futures Modeling System (EFMS) | Canada (CER, 2024) |
| Global Energy and Climate Model (GEC Model) | International Energy Agency (IEA, 2023b) |
| National Energy Modeling System (NEMS) | U.S. Energy Information Administration (EIA, 2024) |
| Hydrogen Delivery Scenario Analysis Model (HDSAM) | Argonne National Laboratory (Elgowainy et al., 2015) |
| Net-Zero Industry Pathways (N-ZIP) Model | UK (DESNZ and BEIS, 2021; Element Energy, 2021) |

**Table 2:** Recent review studies on hydrogen modelling

| Publication | Focus | Coverage |
|---|---|---|
| (Bolat and Thiel, 2014) | bottom-up energy models of hydrogen systems | 10 models |
| (Maryam, 2017) | Modeling the Hydrogen Supply Chain (HSC) problem | |
| (Hanley et al., 2018) | Integrated energy system models | 21 models |
| (Li et al., 2019) | Hydrogen supply chain network (HSCN) design models | 71 publications |
| (Blanco et al., 2022) | Models of hydrogen energy systems | 49 publications |
| (Riera et al., 2023) | Hydrogen production and supply chain modeling | 99 publications |

## 4. Hydrogen modeling elements

This section is divided into seven parts that draws on insights from the methodological review approach elaborated in section 3. This section covers key aspects of the hydrogen supply chain detailing production methods, storage, transportation, trade, demand, hydrogen policies, and modeling strategies. A summary on the current modeling approaches is provided for each sub-section.

*4.1 Hydrogen production sector*

Hydrogen can be produced through several technologies, including steam methane reforming (SMR), electrolysis, gasification (mainly from coal and biogas), biomass conversion, and by-products from industrial processes, as shown in Fig. 3. Currently, SMR is the dominant technology accounting for 49% of the world's hydrogen production today (Riester et al., 2022). It benefits from well-established infrastructure and economies of scale, which results in competitive production costs. Gasification is the second largest technology. However, these production processes will produce carbon dioxide emissions. As the demand for clean energy solutions increases, the scale of electrolysis is expected to expand and account for approximately half of the market share (Statista Research Department, 2024).

Given the environmental impacts of different technologies, we can categorise hydrogen production into different colours, grey, blue, and green, as the three most used concepts. The "grey" designation refers to SMR and gasification processes without CCUS indicating that the hydrogen production process will emit carbon dioxide directly into the atmosphere.

Blue hydrogen, on the other hand, also produces carbon dioxide, but most of these emissions will be stored using CCUS. Blue hydrogen is seen as a transitional technology, offering a lower-carbon alternative to grey hydrogen while renewable energy capacities are being scaled up. Green hydrogen is produced via electrolysis of water, where the electricity used comes from renewable energy sources like wind, solar, or hydroelectric power. This process has zero emissions, as the only by-product is oxygen, and therefore is highly sustainable and is gaining attention as an integral part of future clean energy systems.

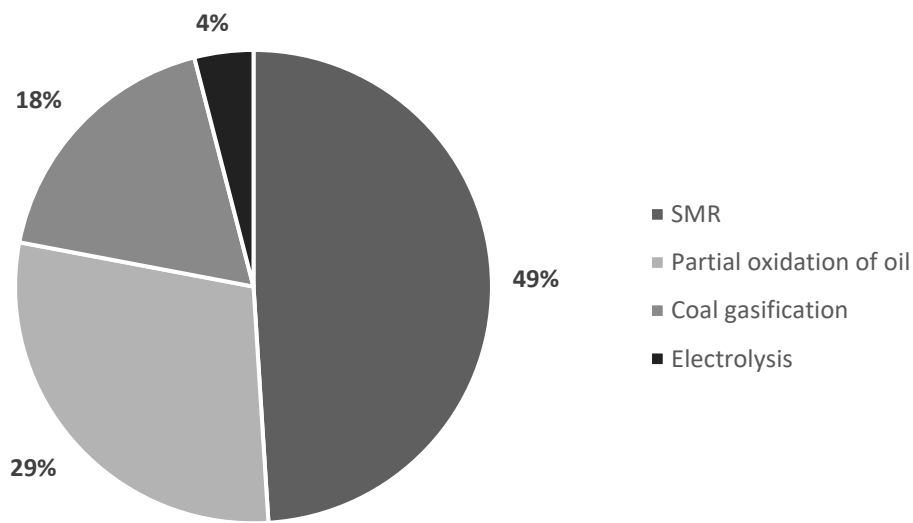

**Fig. 3.** Global shares of hydrogen production technologies
Source: (Riester et al., 2022).

4.1.1 SMR

Steam reforming or steam methane reforming (SMR) is a method for producing syngas (hydrogen and carbon monoxide) by reaction of hydrocarbons with water. Commonly natural gas and water are the feedstock, and hydrogen, carbon monoxide, and carbon dioxide, are the products. The SMR module contains a few steps: firstly, primary reforming consumes carbon compounds in the natural gas, such as methane, and reacts with water, to produce carbon monoxide and hydrogen; Secondly, the water-gas shift reaction happens to convert carbon monoxide and water to carbon dioxide and hydrogen. For the implementation, several pieces of equipment are involved: reactors, separation units, and heat integration unit (IEA, 2019).

The modelling of SMR involves several key aspects, namely, cost, efficiency, and other operating characteristics. In the N-ZIP model[3] the general characteristics of SMR in the operational and planning stage usually involves: CAPEX, fixed OPEX ((£/y)/kW), variable OPEX (£/MWh), overall fuel efficiency (measured by heat value), reference size (MW in final energy consumption), scaling exponent, load factor, and lifetime (year) (Element Energy, 2021). This model can be improved by considering detailed operating characteristics in energy system operations, such as max ramping speed, max on(off) time, and max state change time, in response to the possible inherent failures and feedstock supply uncertainties. In the planning stage, a few more decision variables are involved, such as the year of installation, capacity, location, and operating characteristics above, in typical scenarios.

SMR process is considered by most national energy system modelling practice (DESNZ, 2021). In the UK's national hydrogen strategy, SMR is assumed to be retrofitted with CCUS facilities, and the carbon capture rate is around 90%. A more advanced technology, namely, auto thermal reformation (ATR), is modelled with higher energy conversion and carbon capture efficiencies. Thermal efficiencies of these two technologies are considered around 74% and 86%. In countries with higher ambitions towards green hydrogen, the domestic production of SMR may not be considered. For example, in the IFE-TIMES-Norway model, SMR-produced hydrogen is only modelled as imports (Danebergs et al., 2022). In the energy transition model of the Netherlands, the SMR and ATR are modelled as dispatchable resources rather than a fixed utilisation rate, which offers more flexibility when handling the uncertainties in feedstocks and hydrogen demands (Netherlands 2030, 2023).

4.1.2 Carbon capture and storage (CCS)/CCUS

CCS/CCUS is usually combined with SMR or gasification process in the national energy system modelling. The CCS has three technologies, namely capture at the post-combustion stage, capture at the pre-combustion stage and oxyfuel technology (which consists of burning with oxygen instead of air). CCS consumes electricity, and its performance is measured by captured carbon emissions. Therefore, the features of CCS can also be modelled in terms of cost, efficiency, and other operating or planning characteristics.

---

[3] The N-ZIP model was deployed to generate industry decarbonisation pathways in the UK's industrial decarbonisation strategy and contributed to the analysis in the Climate Change Committee's (CCC) sixth Carbon Budget report (DESNZ and BEIS, 2021; Element Energy, 2020).

The CCS model is an indispensable part of the national hydrogen integration model, but surprisingly, most open-source national energy models, including TIMES and PRIME, don't have built-in CCS modules. In the UK's hydrogen strategy and N-ZIP model, the CCS is located in major industrial clusters (Element Energy, 2020). They are equipped with dedicated pipelines for transporting the captured carbon dioxide to offshore storage. The carbon capture rate is modelled between 95% to 99%, depending on different technology advancement scenarios. In Italy's TEMOA model (Nicoli, 2024), different CCS technologies, including power plant, SMR, gasification, and direct air capture, are modelled separately.

4.1.3 Electrolysis

Electrolysers splits water into its constituent parts—hydrogen and oxygen. There are primarily three types of electrolyser technologies, each differing in their electrolyte and operational characteristics. Firstly, alkaline (AEL) electrolyser operates at relatively low pressures and temperatures and are the most commercially mature technology. It is known for its reliability and robustness but has limitations in dynamic operation and efficiency at higher current densities. Proton exchange membrane (PEM) is commercially available with higher response speeds to the fluctuations of renewables. Solid-oxide electrolyser (SOE) is currently still facing challenges in large-scale deployment due to its higher cost.

In national energy system models, the electrolyser module consumes electricity and water and produces hydrogen and oxygen. The key parameter is efficiency, which is assigned distinguished values in different energy models. In the UK's hydrogen strategy, three technologies of electrolysers are modelled separately, where the energy conversion efficiencies of PEM and AEL are 76% to 82% and 79% to 82%, respectively, evolving with time from 2025 to 2050. Their ability to respond to fluctuating renewables is considered. Some national models model the electrolysers as a general power-to-hydrogen process, such as the Netherland's energy transition model (Netherlands 2030, 2023). In the PRIMES model, the electrolysers are modelled as hydrogen supply and storage, which can be blended with natural gas and smooth the renewable generation curves. In some countries, electrolysis is not considered for its relatively smaller scales compared to fossil fuel-based hydrogen production technologies, such as Italy's Temoa model (Matteo, 2020).

The modelling of electrolysers also covers cost aspects as well. It should be particularly noticed that with the development of offshore wind, significant electrolyser capacities will be installed offshore. This would result in substantial changes to the calculation of costs especially operational expenditures, which is not only related to the installation capacity but also related to the distance to the coast (Michael et al., 2017). These features are rarely reported in the national energy system models where they are usually considered as a fixed value. Another usually omitted aspect of electrolysis is the economic value of byproduct oxygen. In most national energy models, the produced oxygen is not accounted as commodities, but rather released directly into the atmosphere. In some studies, oxygen is modelled as a feedstock to demand sectors such as hospitals and steel manufacturing industries (Song et al., 2021).

4.1.4 Gasification

Compared to SMR, gasification usually converts solid carbon compounds, such as coal, biomass, waste, and water, into carbon monoxide, hydrogen, carbon dioxide, methane, and water vapour. The parameters of the coal gasification process are similar to SMR. Due to the multiple outputs, the fractions of each product in the total final products should be either modelled as fixed parameters or variables (if they can be adjusted according to the technologies). Biomass gasification consumes biomass (including wood, and agricultural residue). The production includes syngas (hydrogen, water, and other hydrocarbons), tar, ashes, and coke (where the latter three are categorised as impurities).

Due to the carbon emission and impurities, the gasification is usually equipped with CCS facilities. In the UK's hydrogen strategy, both coal and biomass gasification are modelled, which provides a lower cost of hydrogen in the initial stage of net zero transition. In the PRIMES model, because gasification is not solely focused on producing hydrogen, but other byproducts as well (such as syngas). Therefore, it is modelled as an energy conversion section with storage capacities, while the byproducts are feedstocks to other sectors. The availability of different feedstocks for gasification, including woody biomass, and black liquor, are modelled separately.

4.1.5 Industrial byproduct

Hydrogen can also be produced as a by-product in several industrial processes, such as the Chlor-Alkali industry, by electrolysis of sodium chloride. Hydrogen is produced as a byproduct at the anode during this process. In the steam cracking of hydrocarbons in ethylene production, hydrogen will also be generated. Compared with other processes above where hydrogen is the major product, these industrial process need to be modelled to consider the purity of the final product, and the energy/cost involved in the separation process. In most of the national energy models, the hydrogen byproducts are neglected. A few practices, such as Temoa in Italy, considered the chemical product of Chlor as a commodity, but the by-product hydrogen is still omitted.

4.1.6 Conversion to hydrogen derivatives

To enhance the utility, ease of transport, and storage, hydrogen can be further converted to various derivatives, especially during long-distance shipping. The hydrogen (gaseous) derivatives mainly include liquid hydrogen, ammonia, methanol ($CH_3OH$), and formic acid. Other less common derivatives include liquid organic hydrogen carriers (LOHC, including methylcyclohexane, dibenzyltoluene, N-ethylcarbazole, toluene, benzene, $H_2O_2$); hydrochloric acid (HCl), hydrogen peroxide ($H_2O_2$), ethanol ($C_2H_5OH$), hydrazine ($N_2H_4$). Techno-economic analysis should be conducted to decide if hydrogen should be converted into these derivatives in different application scenarios. For example, handling liquid or solid derivatives is often considered safer and more manageable than dealing with highly flammable and explosive hydrogen gas, and the energy density will also be higher. Although converting the gaseous hydrogen into liquids, such as ammonia can reduce the transportation cost, the receiving end will need a certain ability to use such derivative directly, such as power generation. Otherwise, if the derivatives need to be converted back to hydrogen, then additional facilities and energy are required, which probably will reduce the cost competitiveness.

Liquid hydrogen is produced by consuming gaseous hydrogen and electricity. It consists of the following subprocesses: 1) compression: first, gaseous hydrogen needs to be compressed to a higher pressure. This is typically achieved through multi-stage compression, with cooling steps in between each stage to remove the heat generated during compression. (2) Cooling: after compression, the high-pressure hydrogen gas needs to be pre-cooled, usually using coolants like helium or nitrogen. (3) Liquefaction: hydrogen is further cooled under low temperatures near its liquefaction point (-253°C) through the Joule-Thomson effect or Claude cycle, ultimately transforming into a liquid state. This step may require several cooling cycles to reach the required low temperature. During the three subprocesses, equipment such as multi-stage compressors, heat exchangers, refrigeration systems, and storage containers are required. To mathematically model this process, economic parameters, including the capital costs, operational costs, maintenance costs, energy consumption and efficiency at each step need to be characterised.

The production of ammonia consumes electricity, hydrogen, and nitrogen. It includes three subprocesses: (1) nitrogen procurement: nitrogen is extracted from the air, usually using an air separation unit (ASU) that employs cryogenic distillation to obtain pure nitrogen. (2) Ammonia synthesis: the prepared hydrogen and nitrogen gases are reacted in a Haber-Bosch synthesis reactor. This process requires conditions of 300 to 550°C and 150 to 300 atmospheres (atm), using an iron-based catalyst. (4) Cooling and separation: the synthesized ammonia gas is cooled and liquefied, while unreacted hydrogen and nitrogen gases are recycled back to the reactor. During these processes, ASU, synthesis tower (reactor), compressors, cooling system, and separation unit are involved, and their economic and efficiency parameters should be characterised in the model.

These hydrogen derivatives are regarded as commodities in national energy models. In the N-ZIP model, the liquefied hydrogen and ammonia are regarded as an industrial process equipped with CCS and modelled as demand for hard-to-electrify sectors such as marine and aviation sectors. In the Netherlands' Energy Transition model, ammonia reforming is regarded as a genetic process and can be flexibly dispatched. In the Balmorel model, the ammonia import, flexible ammonia production process, and cross-border ammonia flow are modelled, and future landscape is predicted (Ravn, 2011; Kountouris et al., 2024). However, the utilisation of ammonia is still at an early stage, most of national energy models such as PRIMES and PyPSA have not incorporated it.

*4.2 Hydrogen storage*

Hydrogen storage is vital for maintaining a stable energy system, especially as we increasingly rely on renewable energy sources. It is essential to consider different time frames for hydrogen storage—such as short-term, seasonal, and long-term—to effectively balance energy supply and demand. Short-term storage helps manage daily energy fluctuations, while seasonal and long-term solutions are crucial for dealing with variations over months or years. By modelling these storage options, we can ensure that energy systems remain reliable and efficient as renewable energy use grows. There are various methods for storing hydrogen on a large scale, each suited for specific conditions and needs. These methods are generally classified into physical and chemical storage approaches, each with its own unique applications and challenges (Osman et al., 2024).

- **Physical storage:** physical storage methods typically involve storing hydrogen as a gas or a liquid, depending on the desired pressure and temperature. One common method is compressed hydrogen storage, where hydrogen is stored under high pressure, often between 200 and 700 bar. This approach is simple and widely used, particularly for short-term storage, but requires heavy, high-pressure tanks due to the large volume of hydrogen, even when compressed.

    Another method, liquid hydrogen storage, involves cooling hydrogen to -253°C to convert it to a liquid, which greatly reduces its volume. While this method allows for more compact storage, it is energy-intensive and prone to losses through evaporation over time. In some cases, underground storage is employed, where hydrogen is stored in large geological formations such as salt caverns, depleted oil and gas fields, or deep aquifers. This method is ideal for long-term and large-scale storage, as these underground structures can hold vast amounts of hydrogen. However, it requires specific geological conditions and significant infrastructure investment. Pipeline storage, also known as linepack, utilizes pipelines to store hydrogen under pressure as it is transported. This method takes advantage of existing infrastructure but is limited by the capacity of the pipelines and the complexity of managing flow rates effectively.

- **Chemical hydrogen storage:** Chemical storage methods involve storing hydrogen in combination with other materials, from which it can be released when needed. One such method is metal hydride storage, where hydrogen reacts with metals to form solid hydrides. This approach allows hydrogen to be stored at lower pressures and offers greater safety, but the materials required are heavy and expensive, making large-scale use less practical. Another option is liquid chemical storage, where hydrogen is stored in liquid compounds such as ammonia or methanol. These liquids have high energy densities and are easier to transport, but separating hydrogen from the liquid requires additional energy and specialized technologies.

A related method is complex hydrides, where hydrogen is stored in more advanced chemical compounds such as borohydrides or aluminium hydrides. These compounds can hold large amounts of hydrogen in a compact form, but the technology to release hydrogen efficiently from them is still under development. Ammonia-based storage is another variation, where hydrogen is stored in the form of ammonia ($NH_3$). Ammonia is easier to store and transport as a liquid under moderate conditions, but converting ammonia back into hydrogen requires energy-intensive processes and advanced technologies. Emerging technologies in nanomaterials and advanced structures are also being explored for hydrogen storage. Materials such as carbon nanotubes and graphene have shown potential for storing hydrogen at the nanoscale, offering the possibility of high storage density. However, these technologies are still in the experimental phase and are not yet widely adopted for practical use.

4.2.1 Hydrogen storage methods

Hydrogen storage plays a crucial role in ensuring flexibility and reliability in renewable-based energy systems, with key methods including underground storage in salt caverns, depleted oil and gas fields, and deep aquifers for long-term use, as well as purpose-built facilities, gas and liquid storage, and advanced solutions like metal hydrides and chemical storage for short-term and specialized applications. Offshore storage solutions, including salt caverns beneath the sea and floating storage systems, also add capacity and flexibility to our energy infrastructure. Integrating various storage methods into energy system planning enables the creation of a robust and adaptable hydrogen energy network that meets future energy demands (Züttel, 2004). This discussion focuses on several prominent storage solutions: salt cavern storage, depleted oil and gas fields, man-made facilities, and linepack systems.

*Salt cavern storage and depleted oil and gas fields*

Hydrogen storage in salt caverns and depleted oil and gas fields presents effective solutions for large-scale and long-term storage of hydrogen. Salt caverns are highly impermeable underground formations that have been utilized for hydrogen storage since the 1970s in the UK and the 1980s in the U.S., achieving an impressive efficiency of around 98%, which enhances the management of injected and extracted hydrogen. These caverns allow for high discharge rates, making them particularly suitable for industrial applications. On the other hand, depleted oil, and gas fields, while typically larger, are more permeable and may contain contaminants that require careful management to ensure hydrogen quality. Historically, water aquifers have also been used to store gas rich in hydrogen, although their maturity and suitability for hydrogen storage remain uncertain.

Salt cavern hydrogen storage is highly cost-effective, with the levelized cost typically estimated to be below €0.6/kg $H_2$, and in some cases, as low as €0.55/kg $H_2$. This makes it one of the most affordable options for hydrogen storage (IEA, 2019). This low operational cost and the economies of scale associated with these storage methods enhance their appeal in the hydrogen economy, leveraging existing geological formations and infrastructure to provide a cost-effective means of managing hydrogen as a vital energy resource. Notably, the efficiency of hydrogen storage in these systems typically falls between 85% and 95%, factoring in losses during the charging and discharging cycles. These underground storage methods can support significant capacities, often exceeding 100 MWh, depending on the geological characteristics and advancements in storage technology. Proper management of hydrogen flow rates during charging and discharging is crucial for balancing energy supply and demand effectively. It is also essential to ensure that the geological conditions are suitable for safe and efficient hydrogen storage, as not every location possesses the requisite qualities for this purpose (Muhammed et al., 2022).

*Man-Made Storage Facilities*

Custom-built storage facilities offer customized solutions for hydrogen storage, accommodating a range of operational capacities and functionalities. These facilities can be categorized into two main types: onshore storage systems and above-ground storage tanks. Custom-built hydrogen storage systems require significant initial capital investments, which can vary widely based on factors such as the materials chosen for construction and the technology employed. The financial implications for operational and maintenance costs are also substantial, encompassing labor, maintenance, and energy requirements. These considerations underline the economic challenges associated with developing efficient hydrogen storage solutions (Papadias et al., 2021).

Recently, models were developed to consider hydrogen storage above-ground in tanks and in underground caverns. Above-ground tank storage is generally appropriate for short-duration applications (typically no more than a few days), whereas underground storage is optimized for longer seasonal durations (120 to 150 days). Above-ground storage tanks are often said to cost approximately €380-580/kg $H_2$, more expensive than the comparatively low cost of going underground. Nonetheless, the contribution of tank storage to the overall levelized cost of hydrogen dispensed is typically small at around €0.16/kg $H_2$, while additional compression costs contribute approximately €0.39/kg $H_2$. On the other hand, the levelized cost of liquid hydrogen storage for a 14-day period, including liquefaction, is approximately €1.28/kg $H_2$. These results suggest that storage provision has a relatively low contribution to the system for short-term applications, while long-duration, seasonal storage has a higher cost (Burke et al., 2024).

Onshore hydrogen storage is essential for promising cost-effective hydrogen operations and plays an important role in commercializing hydrogen fuel. Such facilities can include different storage technologies like pressurized tanks and metal hydrides to achieve high efficiency at low cost. Their modular design enables quick response to fluctuations in hydrogen demand, making them well-suited for renewable energy integration.

Both onshore storage facilities need to consider several parameters for their modelling, namely capacity, efficiency, and operational dynamics. Simulations are often used to help assess flow rates, safety measures, and geological properties of the sites used for storage. Using this integrated method allows for optimization across both the design and operation of the hydrogen storage system while addressing the economic and environmental aspects of its implementation (Mulky et al., 2024).

*Linepack Systems*

Linepack systems utilize existing gas pipelines to store hydrogen under pressure, enabling efficient integration with transportation networks. These systems typically incur lower costs compared to other storage methods, as they leverage current infrastructure. The potential storage capacity is influenced by the size and layout of the pipelines. For example, in larger pipelines, it is feasible to store between 100 to 300 tons of hydrogen, with a levelized cost estimated at approximately €0.049/kg $H_2$ in 36- and 48-inch diameter pipelines of 100 km length. For a 12-inch pipeline, the cost is approximately €0.28/kg $H_2$, decreasing further to €0.078 and €0.049 per kilogram for 24-inch and 36-inch pipelines, respectively (Burke et al., 2024). This method is particularly effective for very short storage durations, typically one day or less. However, evaluating the performance of linepack systems can be complex, as it necessitates comprehensive analysis and often requires additional modelling to accurately capture the dynamic fluctuations of supply and demand.

Choosing the appropriate hydrogen storage solution depends on a combination of economic viability, geographical context, and energy demands. Salt cavern and depleted oil and gas field storage methods provide substantial capacity for long-term storage but are limited by geological factors. In contrast, man-made facilities offer customized solutions, while linepack systems leverage existing infrastructure for efficient energy management. As shown in Table 3, the linepack system provides a short-duration storage option (typically one day or less) with a capacity of 100–300 tonnes of $H_2$ at a relatively low levelized storage cost of €0.05/kg $H_2$ or less. Continued advancements in technology and materials will likely enhance the efficiency and reduce costs across these storage methods, further facilitating the growth of a hydrogen-driven energy future (Wang et al., 2023).

Table 3: Overview of hydrogen storage solutions and their associated costs.

| Storage technology | Typical duration of storage | Tonnes of H2 typically stored | Levelized storage cost range (€/kg H2) |
|---|---|---|---|
| Linepack system | 1 day or less | 100-300 | 0.048 or less |
| Salt cavern | 2-4 months | 500-1000 | 0.57-1.15 |
| Above ground pressurized tank, GH2 | 1-2 days | 0.3-1 | 0.28-0.48 |
| Above ground liquid tank, LH2 | 1-2 weeks | 5-10 | 0.057-0.115 |

Source: (Burke et al., 2024)

Hydrogen storage modelling encompasses a wide range of approaches, as illustrated by several prominent models detailed in Table 4, each tailored to specific applications and regions. Optimization models like TIMES and MARKAL focus on long-term energy system planning and policy analysis, with an emphasis on storage types such as underground caverns and chemical storage. These models face challenges such as the complexity of integrating various energy sources and future technology predictions. Simulation models like EnergyPLAN and NEMS, on the other hand, simulate national energy policies, with difficulties arising from regional variations and sensitivity to input assumptions. The Hydrogen Delivery Scenario Analysis Model (HDSAM) highlights infrastructure modelling, addressing the logistics of hydrogen storage and transportation. Overall, these models are essential for energy planning, but they require extensive data integration and validation to ensure accuracy in addressing hydrogen storage's technical and logistical challenges.

**Table 4:** Overview of hydrogen storage modeling

| Model Family / Custom Model | Country/Organization | Model Type | Open source | Objective | Key Features | Application Focus | Storage Types Modeled | Technical Challenges | Modeling Domain | References |
|---|---|---|---|---|---|---|---|---|---|---|
| TIMES (The Integrated MARKAL-EFOM System) | Multiple (Norway, Portugal, etc.) | Opt. | Semi | Cost minimization, emissions reduction | Integrated assessment of energy systems, including hydrogen storage | Long-term energy planning | Underground (Salt Caverns, Depleted Oil Fields), Compressed Gas | Complexity in modeling interactions between various energy sources and storage methods; requires comprehensive data integration. | Energy systems modeling | Hoffmann, 2024; Pedersen, 2021 |
| MARKAL (Market and Allocation) | China | Opt. | Semi | | Focus on technology-rich energy system modeling | Policy analysis and scenario planning | Various including Chemical Storage | Difficulty in accurately predicting future technology advancements and their impacts on hydrogen storage scenarios. | Energy policy modeling | Hoffmann, 2024; Pedersen, 2021 |
| PyPSA (Python for Power System Analysis) | Poland | Opt. | Yes | Cost minimization | Power system optimization with a focus on renewable integration | Short-term and long-term planning | Compressed Gas, Liquid Hydrogen | Challenges in real-time data management and flow optimization across complex networks; high computational demands. | Power systems analysis | Hoffmann, 2024; Usman, 2022 |
| Balmorel | Poland | Opt. | Yes | Economic optimization of electricity markets | Economic modeling of electricity markets | Market analysis | Various including Underground Storage | Limited ability to capture dynamic market behaviours and regulatory changes affecting hydrogen pricing and demand. | Economic modeling | Hoffmann, 2024; Usman, 2022 |
| PRIMES Model | EU Commission, Poland | Opt. | No | Cost minimization, policy analysis, | Model for energy system analysis in | EU energy policy | Compressed Gas, Chemical Storage | Integration of diverse energy policies and regulations across multiple | Energy policy analysis | Mulky, 2024; Usman, 2022 |

| Model | Region | Type | Hydrogen | Focus | Description | Application | Storage Type | Challenges | Key Area | Reference |
|---|---|---|---|---|---|---|---|---|---|---|
| | | | | emissions reduction | the EU context | | | countries adds complexity to modeling efforts. | | |
| EnergyPLAN | Spain | Sim. | Yes | Policy and renewable energy scenario analysis | Detailed simulation of energy systems with a focus on renewables | Energy transition strategies | Underground, Compressed Gas | High sensitivity to input assumptions; requires robust validation against real-world data for accuracy. | Renewable energy systems | Pedersen, 2021; Usman, 2022 |
| Canada Energy Futures Modeling System (EFMS) | Canada | Sim. | No | Energy and climate policy assessment | Comprehensive modeling of Canadian energy systems | National energy policy | Various including Underground Storage | Variability in renewable energy sources complicates long-term forecasting and planning for hydrogen integration. | National energy systems | CER, 2024 |
| Global Energy and Climate Model (GEC Model) | International Energy Agency | Sim. | No | Energy and climate policy assessment | Global assessment of energy and climate policies | Climate change mitigation | Various including Chemical Storage | Challenges in aligning global models with local conditions; requires extensive data collection from diverse regions. | Global climate modeling | IEA, 2023b |
| National Energy Modeling System (NEMS) | U.S. Energy Information Administration | Sim. | Yes | Energy and climate policy assessment | Detailed modeling of U.S. energy markets | National energy policy | Compressed Gas, Chemical Storage | Difficulty in capturing regional variations in energy consumption patterns and infrastructure capabilities. | National energy analysis | Mulky, 2024; Usman, 2022 |
| Hydrogen Delivery Scenario Analysis Model (HDSAM) | Argonne National Laboratory | Sim. | Yes | Hydrogen infrastructure and logistics scenario analysis | Focused on hydrogen delivery infrastructure and logistics | Hydrogen economy development | Pipeline Storage (Linepack), Compressed Gas | Complexity in modeling logistics networks; requires precise data on transportation | Hydrogen infrastructure modeling | Mulky, 2024; Argonne, 2007 |

| | efficiencies and costs. |
|---|---|
| Notes: Opt: Optimization; Sim: Simulation, Semi: Semi-open source | |

### 4.2.2 Carbon storages

Carbon storage plays a crucial role in hydrogen production pathways, particularly in blue hydrogen, where carbon dioxide ($CO_2$) emissions from steam methane reforming (SMR) or autothermal reforming (ATR) are captured and stored instead of being released into the atmosphere. This process enhances the sustainability of hydrogen production by reducing its carbon footprint. Various carbon storage methods exist, including geological storage, ocean storage, mineral carbonation, and biological sequestration. Among these, geological storage is the most widely implemented, involving the injection of $CO_2$ into depleted oil and gas reservoirs, deep saline aquifers, or unmineable coal seams. These formations provide secure, long-term containment due to their impermeable rock layers, which act as natural barriers against leakage. Integrating effective carbon storage solutions with hydrogen production can support large-scale decarbonization efforts and the transition to a low-emission energy system (Gayathri et al., 2021).

Geological storage of carbon is particularly noteworthy for its potential to store vast amounts of $CO_2$ safely and effectively. Research has indicated that depleted oil and gas fields are well-suited for this purpose, as they have previously contained hydrocarbons and possess the necessary geological characteristics for secure $CO_2$ storage. Saline aquifers, which are permeable rock formations filled with saltwater, also present significant potential for carbon sequestration due to their extensive availability and capacity. Effective monitoring and verification systems are essential to ensure the integrity of geological storage sites and to mitigate any potential environmental impacts associated with leakage (Fajardy et al., 2021).

In the context of hydrogen storage, advancements in technology and materials are becoming increasingly relevant. Hydrogen storage systems are evolving to complement carbon storage methods, particularly in integrated energy systems that prioritize renewable energy sources. For example, the co-utilization of hydrogen and carbon storage can facilitate the development of low-carbon hydrogen, including green hydrogen, which is produced through electrolysis powered by renewable energy, and blue hydrogen, where carbon capture and storage (CCS) is used to reduce $CO_2$ emissions from hydrogen production processes such as steam methane reforming (SMR) or autothermal reforming (ATR). By integrating these approaches, a more sustainable hydrogen economy can be established, where $CO_2$ emissions are effectively managed while ensuring a reliable supply of clean energy.

As both hydrogen and carbon storage technologies continue to develop, ongoing research is necessary to optimize their integration within energy systems. This includes evaluating the potential synergies between different storage methods, assessing the economic viability of combined approaches, and exploring innovative materials for more efficient storage solutions. The future of sustainable energy will likely depend on a holistic approach that incorporates multiple storage strategies, ensuring a resilient and environmentally responsible energy infrastructure.

*4.3 Hydrogen transportation*

Hydrogen can be transported using a variety of approaches such as compressed gas cylinders, hydrogen carriers, pipeline transportation, cryogenic liquid tankers (Faye et al., 2022; Raj et al., 2024). The goal is to select a transport option that can deliver a higher volume of hydrogen in a financially feasible and efficient manner. At ambient temperatures, hydrogen is a gas with a very low density of 0.089 kg/m$^3$ (Züttel, 2003). Liquified hydrogen can be achieved at lower temperatures (-253°C), which increases the density to 70.8 kg/m$^3$. Hydrogen is in a solid form at -262°C with a density of 70.6 kg/m$^3$.

- **Pipeline transportation:** This option entails delivering hydrogen using dedicated pipelines or repurposing existing natural gas or oil pipeline infrastructure. The advantage of pipeline transportation is the ability to cost-effectively transport large volumes of hydrogen over long distances (Faye et al., 2022; Raj et al., 2024). The upfront costs for pipeline infrastructure is significant, however once completed, the operating costs are expected to remain small (Faye et al., 2022). To circumvent large upfront capital costs for dedicated hydrogen pipeline, gas operators are considering the possibility of blending hydrogen into existing gas pipelines. Recent studies have demonstrated that the direct blending of green hydrogen of up to 20% into the gas network is feasible without any gas network issues, negating the need to upgrade the gas network infrastructure (Ekhtiari et al., 2022, 2020). Developing 100% hydrogen networks is a key step toward decarbonizing energy systems, ensuring hydrogen's role as a sustainable alternative to natural gas. However, transitioning to a fully hydrogen-powered network requires strong political commitment, detailed strategic planning, and expedited approvals for large-scale energy infrastructure. Several countries have set ambitious targets for hydrogen deployment, including plans to develop large-scale hydrogen production capacities of around 10GW by 2030 (e.g., UK) alongside dedicated hydrogen transmission infrastructure (Hydrogen UK, 2023). Projects like FutureGrid and HyDeploy are assessing the feasibility of transporting pure hydrogen and blending it with natural gas within existing transmission networks, paving the way for a hydrogen-based energy future.

- **Compressed gas cylinder:** Transporting hydrogen using compressed gas cylinders is ideal for short distances. Hydrogen can be stored in gas cylinders and transported on land via trailers. The maximum volume that can be transported on trailers depends on the height of individual cylinders or tubes (Faye et al., 2022). Current commercial pressure vessels are divided into five categories (type I – V) depending on their composite content (Daghia et al., 2020). Type I pressure vessels are made of metal such as steel and can transport up to 250 kg H$_2$ at 200 bar pressures (Reddi et al., 2018). While Type III and IV can transport up to 1000 kg $H_2$ at higher pressures (500 bar). Type V pressure vessel is a new technology and in the development stage (Reddi et al., 2018; Daghia et al., 2020).

- **Cryogenic liquid tankers:** hydrogen can also be transported in liquid form via ships, trailers, or railway. This entails cooling hydrogen down to around -253°C in a process known as liquefaction which increases the volumetric density of hydrogen. However, liquefaction is an energy-intensive recording approximately 40% energy losses (Aziz, 2021; Faye et al., 2022) Cryogenic tankers are typically deployed for medium to long-term transportation (Hosseini, 2023).

- **Hydrogen carriers:** Beyond pipeline and gaseous transport options, hydrogen can also be stored and transported using chemical compounds known as hydrogen carriers, including ammonia, methane, and liquid organic hydrogen carriers (LOHCs). These carriers help overcome the challenges associated with conventional gaseous and liquid hydrogen transport methods, such as high liquefaction costs, energy consumption, and boil-off losses (Papadias et al., 2021). The general transport chain for liquid hydrogen carriers typically involves: (1) transporting the liquid carrier from the production facility to a storage terminal via rail, (2) transferring it to a decomposition plant via trucks, and (3) delivering the extracted gaseous hydrogen to end users via additional transport infrastructure (Papadias et al., 2021).

Among these carriers, ammonia has emerged as a promising hydrogen vector due to its relatively high hydrogen content and established global transport infrastructure. Currently, ammonia is predominantly used as a feedstock for fertilizers and various industrial applications, such as plastics, explosives, and synthetic fibres (Aziz, 2021; IEA, 2021b; Papadias et al., 2021). It can be synthesized from both fossil fuel-based and renewable sources, with approximately 70% of global ammonia production relying on natural gas through steam-methane reforming, while the remainder is derived from coal gasification (IEA, 2021b). In comparison to other carriers, ammonia has a higher boiling point (-33 °C), which requires less energy when converting to and preserving in liquid form (Patonia and Poudineh, 2022) (see Table 5). In addition, the higher boiling point ensures a lower thermodynamic loss when stored and transported. Ammonia can be transported using pipelines or low-pressure tanks. Crucially, ammonia also has a higher volumetric hydrogen content per cubic meter compared to other energy carriers.

Another important carrier is methanol, which is synthesized through syngas production using fossil fuel feedstocks such as natural gas and coal. Methanol is widely utilized as an industrial feedstock in chemical manufacturing, and its production costs are largely driven by syngas generation, which accounts for approximately 60% of the capital costs of a methanol plant (Papadias et al., 2021). As shown in Table 5, methanol has the second-highest volumetric hydrogen content among hydrogen carriers, making it a viable alternative for large-scale hydrogen transport and storage.

Liquid Organic Hydrogen Carriers (LOHCs) represent another potential solution for hydrogen storage and transportation. LOHCs are organic compounds that remain liquid at ambient conditions and can undergo reversible hydrogenation and dehydrogenation processes at delivery points. These carriers offer significant advantages, including lower energy losses during storage, compatibility with existing fuel infrastructure, and the ability to release high-purity hydrogen upon dehydrogenation (Aakko-Saksa et al., 2018; Patonia and Poudineh, 2022; Chu et al., 2023). The candidates for LCOHs are broad and include toluene/methylcyclohexane (MCH), naphthalene-decalin, benzene-cyclohexane and biphenyl/bicyclohexyl among others. Toulene is considered the most established in the market (Patonia and Poudineh, 2022).

**Table 5:** Characteristics of selected hydrogen carriers.

| Characteristics | Liquid hydrogen | Ammonia | Methanol | Methylcyclohexane (MCH) |
|---|---|---|---|---|
| Density under normal conditions (kg/m$^3$) | 0.08375 | 0.73 | 791.4 | 866.9 |
| Melting point (ºC) | -259.16 | -77.73 | -97.6 | -126.3 |
| Boiling point (ºC) | -252.87 | -33.34 | 64.7 | 101 |
| Volumetric energy density (Wh/L) | 8.49 | 12.92-14.4 | 11.40-11.88 | 5.66 |
| Volumetric H$_2$ content (kgH$_2$/m$^3$) | 70.8 | 107.7-120 | 95.04-99 | 47.1 |
| Common production process | Electrolysis, SMR, coal gasification | Haber-Bosch process | Carbon hydrogenation/ Methanation | Hydrogenation of toluene |

Source: Adapted from (Patonia and Poudineh, 2022)

Table 6 compares the estimated minimum levelized costs for key stages in the value chains of liquid hydrogen, ammonia, methylcyclohexane (MCH), and methanol, expressed in €/kg H$_2$. Liquid hydrogen has the highest total cost due to energy-intensive liquefaction and storage. Ammonia appears more cost-effective in storage and transport compared to liquid hydrogen, while MCH has moderate costs but requires additional processing for hydrogen extraction. Methanol shows relatively lower costs for conversion and transport but includes carbon capture and storage (CCS) expenses. This comparison helps assess the economic feasibility of different hydrogen carriers for large-scale energy transport.

**Table 6:** Estimated minimum levelized costs for the key stages in the value chains of liquid hydrogen, ammonia, MCH, and methanol (€/kg $H_2$)

| Focus fuel | Production | Conversion | | Storage | Shipping | Reconversion | Sum of the components |
|---|---|---|---|---|---|---|---|
| **Liquid hydrogen** | >0.96 | Liquefaction | 1.6-3.4 | >4.56 | 1.6-2.5 | N/A | >8.62 |
| **Ammonia** | >2.11 | | 0.72-1.44 | >0.48 | 0.53-0.78 | 0.28-1.53 | >4.14 |
| **Methylcyclohexane (MCH)** | >1.29 | N/A | | | 1.31-1.99 | 0.51-1.17 | >3.13 |
| **Methanol** | >1.17 | | | | 0.65-0.83 | 0.41-1.07 (dehydrogenation) + >0.6 (CCS) | >2.81 |

Source: Adapted from (Patonia and Poudineh, 2022)

The reviewed energy system models adopt a variety of approaches to handle hydrogen transport, as indicated in Table 7. The majority of these models incorporate pipelines to transport hydrogen within and between geographical jurisdictions. Recognizing that a purpose-built hydrogen infrastructure network will take time to materialize, the models allow for the repurposing of existing natural gas infrastructure, specifically methane pipelines. This approach is considered a cost-effective short-term solution, enabling the transition to hydrogen without requiring immediate large-scale investment in new infrastructure. This finding is consistent with (Zhang et al., 2025), who discovered that most studies adopt pipeline models to evaluate whether existing or dedicated transport infrastructure can accommodate either hydrogen injection or full-scale hydrogen transportation. However, the feasibility of repurposing natural gas pipelines depends on several factors, including material compatibility, hydrogen embrittlement risks, pressure requirements, and volumetric capacity constraints due to hydrogen's lower density compared to methane.

For example, in the PyPSA model, existing methane pipelines can be decommissioned and repurposed for hydrogen transport. The overnight investment cost of a repurposed hydrogen pipeline is estimated at €105.88 /MW/km, significantly lower than the €226.47 /MW/km required for a dedicated hydrogen pipeline (Neumann et al., 2023). While repurposed pipelines provide a transitional solution, the long-term deployment of dedicated hydrogen pipelines may be necessary to ensure system efficiency and safety.

In terms of hydrogen carriers, models such as PyPSA and IEA Global Energy and Climate Model (GEC) produce and transport synthetic fuels such as ammonia and methane. While models such as Balmorel and PRIMES are limited to the production of synthetic fuels. For instance, the IEA Global Energy and Climate Model (GEC) has the option of meeting domestic hydrogen demand with cheaper imported liquid hydrogen-based fuels. Hydrogen-based fuel such as ammonia is imported via ship and dehydrogenated at the delivery point. The model then factors in the cost of conversion and transportation in the overall hydrogen cost optimization framework. Furthermore, detailed hydrogen transport assessments are largely confined to specialised tools. One such model is the Hydrogen Delivery Scenario Analysis Model (HDSAM) which is an Excel-based engineering-economic model that computes the cost of delivering hydrogen from a production facility to a delivery point (Elgowainy et al., 2015). The HDSAM incorporates detailed configurations covering hydrogen vehicle type, transmission, and distribution modes (i,e., liquid truck, pipeline, tube-trailer), and refuelling station parameters.

**Table 7:** Comparison of hydrogen transportation in selected energy system models

| Model | Hydrogen transportation | Special model features | Sources |
|---|---|---|---|
| TIMES | <ul><li>Hydrogen pipelines,</li><li>Road transport (tankers, tube-trailers)</li><li>Ship transport</li></ul> | Model specific | (Kannan and Turton, 2014; Balyk et al., 2022; Danebergs et al., 2022; Laguna et al., 2022) |
| PyPSA | <ul><li>Repurposed hydrogen pipelines</li><li>Dedicated hydrogen pipelines</li><li>Submarine pipelines</li><li>Shipping of methanol</li></ul> | <ul><li>Decommissioning of gas pipelines</li></ul> | (Neumann et al., 2023) |
| Balmorel | <ul><li>Repurposed natural gas pipelines</li><li>New hydrogen pipelines</li></ul> | <ul><li>Model is allowed to invest in new hydrogen infrastructure.</li><li>Hydrogen imports from nations outside the spatial scope</li></ul> | (Kountouris et al., 2024) |
| PRIMES Model | Hydrogen pipelines | <ul><li>Dedicated pipelines and hydrogen blending of up to 15%</li><li>Model allows the use of hydrogen carriers such as methane and synthetic liquid fuels</li></ul> | (E3M, 2018) |
| EnergyPLAN | No hydrogen transport method | n/a | (Lund and Thellufsen, 2024) |
| Canada Energy Futures Modeling Framework | n/a | n/a | |
| Global Energy and Climate Model (GEC Model) | <ul><li>Pipeline transport for gaseous hydrogen</li><li>Ship transport for liquid hydrogen and hydrogen carriers (i.e., ammonia and synthetic liquid hydrocarbons)</li></ul> | <ul><li>Demand for gaseous hydrogen can be satisfied with liquid hydrogen imports</li><li>Costs of hydrogen transport and conversion taken into consideration</li></ul> | (IEA, 2023b) |

| EIA National Energy Modeling System (NEMS) | Hydrogen pipelines | • Transmission modeling within and between census regions<br>• Easy expansion of transmission system | (EIA, 2024) |
|---|---|---|---|
| Hydrogen Delivery Scenario Analysis Model (HDSAM) | • Dedicated hydrogen pipelines<br>• Tube trailer<br>• Liquid $H_2$ truck | • Distinguishes between transmission and distribution modes | (Elgowainy et al., 2015) |

*4.4 Hydrogen trade*

The source of hydrogen comes either from international import or domestic production. As a fuel energy carrier like natural gas, the global enthusiasm, and ambitions for net zero stimulate considerable hydrogen demand and production, as well as its derivatives. Due to the imbalanced nature of endowment and manufacturing capabilities, the local supply-demand balances are not uniformly distributed across the globe. For instance, countries with vast solar fields or wind farms (e.g., parts of the Middle East, North Africa, and some regions in Australia) can potentially produce large amounts of green hydrogen more economically than regions with limited renewable resources. This imbalance will lead to a substantial international hydrogen market in the future. As projected by McKinsey, the cross-border hydrogen trading potential could reach 200 Mt per year (McKinsey & Company, 2023). An import model helps determine the logistics, costs, and infrastructure needed to transport the hydrogen to meet domestic needs.

Modelling the interflows among countries usually requires an additional toolbox which is typically not included in one country's energy system model. This model is usually used to identify investment opportunities and help to allocate capital globally, and set the boundary conditions, such as availability and prices, for national energy balance analysis. Compared with the national energy model with a clear objective and timeframe, the international trading model is still at an early stage. To identify international transaction opportunities, the model is usually established from economic optimization perspectives, considering demand forecast, supply potential, production cost, transportation cost/limits, production limits, and other diversified constraints such as national energy security constraints (McKinsey & Company, 2022). Conversely, the production cost and limits and demand forecast depend on the national energy system modelling results. Therefore, these two toolboxes should work collaboratively to achieve comprehensive analysis results.

The hydrogen import and export modelling has attracted attention in resource-abundant countries and regions with well-developed hydrogen infrastructures, such as Australia, EU (Oxford Institute for Energy Studies, 2023), the UK (Hydrogen UK, 2024), Ireland, as well as other countries with clear hydrogen vision, such as Germany, Belgium, and Japan. In Australia, South Australia's hydrogen export modelling tool was released to assess hydrogen export opportunities and investments (Government of South Australia, 2020) (but is no longer publicly available). The exporting timeline, max volume, hydrogen type, electricity price, etc., can be characterised as parameters to assess the investment-return report. Another online modelling tool, AusH2 (Australian Government, 2024) is still available with limited functions to show the potential hydrogen export locations. In Ireland, AFRY conducted a comprehensive evaluation on Ireland's hydrogen export potential covering the levelized hydrogen production cost from offshore wind, logistic costs (pipelines and ships), and hydrogen receiving capability of cross-Europe countries (AFRY, 2023).

In national energy models, the hydrogen import can be modelled in similar ways as other fuels, characterised by time-referenced qualities with constraints and price. However, it is rarely considered in the current modelling framework, including typical model structure IEA's TIMES model (Richard, 2016), and JRC-EU-TIMES model. As hydrogen can be modelled as commodities in the TIMES model, it is relatively easy to add an additional hydrogen import based on the international trading analysis results, in addition to the domestic hydrogen production. Additional links to subsequential processes, such as storage and transportation, as well as other energy consumption sectors, need to be established.

*4.5 Hydrogen demand*

In the latest global hydrogen review, the IEA estimates global hydrogen demand at more than 97 Mt in 2023 and is projected to reach approximately 100 Mt in 2024. The bulk of current hydrogen demand is largely confined to traditional applications such as refining, chemical production, and steel manufacturing. The various reviewed models adopt a variety of approaches to model end-use demand for hydrogen. Table 8 summarizes the approaches to model end-use demand. In most cases, the common approach is to disaggregate hydrogen demand by end-use sector. The hydrogen demand is generally then provided either exogenously as a fixed time-series input such as in the PyPSA model or estimated endogenously within the model framework and then fed into a hydrogen module. For example, in the U.S. EIA NEMS model, projected hydrogen demand is estimated from four separate modules, namely, the industry, liquid fuels market, transport, and electricity. The demand aggregated by census region on an annual or seasonal basis and fed into the hydrogen market module (HMM). The EnergyPLAN model estimates hourly balances of hydrogen which is then used to meet demands from industry, transport, micro combined heat and power (CHP), boilers and power plants (Lund and Thellufsen, 2024).

*4.6 Spatial and temporal resolutions*

Most of the reviewed models adopt higher temporal resolutions (i.e., hourly) for selected input data (e.g. demand) and hydrogen assessments (see Table 8). This is essential since a lower temporal resolution is insufficient to evaluate the flexibility of electrolysers and assess the integration of hydrogen in an energy system (Blanco et al., 2022). The main challenge for the models is to weigh-up the trade-offs between richer temporal insights and high computational burden (Langer et al., 2024; Zhang et al., 2025). As a result, models have resorted to adopting representative time-slices or seasonal profiles (e.g. EIA NEMS model). The most flexible model appears to be the EIA NEMS model which proposes a mixture of annual and seasonal timescales for hydrogen demand profiles. The spatial coverage of the reviewed models is often large, encompassing several internal or external regions.

**Table 8:** Hydrogen demand modeling across selected energy system models.

| Model | Sector | | | | | Modeling aspects | | |
|---|---|---|---|---|---|---|---|---|
| | Electricity | Industry | Residential | Transportation | Refining | Exogenous/Endogenous demand | Spatial resolution | Temporal resolution |
| Balmorel | ✓ | ✓ | × | ✓ | ✓ | • Exogenous direct hydrogen demand<br>• Exogenous synthetic fuel demand | EU 27, the United Kingdom, Norway, Switzerland, and Balkan nations | |
| Canada Energy Futures Modeling Framework | ✓ | ✓ | ✓ | ✓ | × | Not stated | 10 provinces and three territories | Not stated |
| EnergyPLAN | ✓ | ✓ | ✓ | ✓ | ✓ | Exogenous | Customizable | Hourly |
| HDSAM | × | × | × | ✓ | × | Exogenous | Representative urban, rural interstate, or combined urban and rural markets | Hourly for one day |
| IEA GEC Model | ✓ | × | ✓ | × | ✓ | Endogenous | 26 regions | Hourly |
| PRIMES | n/a | ✓ | ✓ | ✓ | n/a | Endogenous | EU-28 | Not stated |
| PyPSA | × | ✓ | × | × | ✓ | Exogenous | | |
| TIMES | | | | | | | Country dependent | |
| EIA National Energy Modeling System (NEMS) | ✓ | ✓ | × | ✓ | ✓ | Endogenous | nine census divisions | Seasonal for electricity sector and annual for all others |

Source: (Elgowainy et al., 2015; E3M, 2018; IEA, 2023b; CER, 2024; EIA, 2024; Lund and Thellufsen, 2024; Kountouris et al., 2024).

*4.7 Hydrogen policies*

The development of the hydrogen sector will entail significant government supports on both the supply and demand sides. Direct grants are the most popular mechanism applied in developed economies to reduce capital costs and derisking investments (IEA, 2024). Whereas, emerging market and developing economies (EMDEs) prefer tax credits to stimulate the hydrogen sector. Hence, it is critical for models to account for domestic hydrogen policies that can influence the development of the sector. Most of the reviewed models' factor in hydrogen policies in their modeling framework. Typically, existing models deploy policies in two different approaches. The first is approach is to define an exogenous tax or policy parameter. The second approach is to define a policy scenario. The Balmorel model adopts both approaches, whereby different policies can be modelled with a scenario policy-add on or a clearly defined parameter. The policy add-ons can be applied to specific regions, technologies, and time periods.

In 2023, the U.S. introduced the Inflation Reduction Act (IRA) which incorporates two tax credits for the hydrogen sector as elaborated earlier (45V and 45Q). The EIA NEMS hydrogen market module (HMM) factors in both tax credits to evaluate the impact of the credit on the economic competitiveness of hydrogen production. In the case of the production tax credit (45V), the model assumes only hydrogen production from electrolysis are eligible to receive the credits (see Table 9) (EIA, 2024). Whereas, in the case of the carbon sequestration tax credit (45Q), all CCS technologies are assumed to receive the credits instead of the 45V. Similarly, the PRIMES model adopts hydrogen infrastructure policies that encompass hydrogen transport and distribution. Likewise, the IEA GEC model incorporates policy constraints that considers hydrogen deployment targets, $CO_2$ prices, and regional policy packages.

**Table 9:** Modeling hydrogen policies

| Model | Hydrogen policies | Details | Source |
|---|---|---|---|
| PyPSA | Not stated | Not stated | (Neumann et al., 2023) |
| Balmorel | Yes | • Representation of taxes and subsidies<br>• Flexibility to add additional policy measures with policy add-ons | (Ea Energianalyse, 2018) |
| PRIMES Model | Yes | • Hydrogen infrastructure policies covering transport and distribution. | (E3M, 2018) |
| EnergyPLAN | No | n/a | (Lund and Thellufsen, 2024) |
| Canada Energy Futures Model | n/a | Not stated | (CER, 2024) |
| Global Energy and Climate Model (GEC Model) | Yes | • Inclusion of policy constraints such as $CO_2$ prices or hydrogen deployment targets.<br>• National/Regional energy policies: IRA, Fit for 55 (EU), Climate Change Bill (Australia), GX Green Transformation (Japan) | (IEA, 2023b) |
| EIA National Energy Modeling System (NEMS) | Yes | • IRA 45Q and 45V production tax credit<br>• Assume all CCS technologies opt for the 45Q credits instead of the 45V | (EIA, 2024) |
| Hydrogen Delivery Scenario Analysis Model (HDSAM) | No | n/a | (Elgowainy et al., 2015) |

Note: TIMES models are highly country dependent.

## 5. Suggestions for policy modeling

This section is divided into six parts and provides modeling suggestions for key components of the hydrogen supply chain based on insights presented in section 4.

*5.1 Suggestions for production*

(1) Diversified hydrogen (and the derivatives) production routes are suggested to be incorporated into the national energy model to identify the best production option, including various production technologies (such as gasification and autothermal reforming) and energy carriers (such as ammonia and methanol). Ideally, this should be in in line with national policy priorities. The current national energy system models usually include the two most common hydrogen production technologies, namely, SMR and electrolysers only. Although these two technologies account for more than half of the global hydrogen production, they may not be the most economic or in the case of SMR environmental-friendly options given the various natural endowments of different countries. Moreover, hydrogen derivatives, such as ammonia and methanol, are usually not considered separately in the national model. These shortcomings in the model will affect the generality and global optimality in finding the best hydrogen integration roadmap.

(2) Richer technology features should be incorporated into the model to simulate high-resolution operation for better accuracy. The macroscopic national energy system model, such as TIMES, is usually formulated based on long-term energy supply-demand balance, such as days and years. While this significantly reduces the complexity of the model, it will fail to capture some important features in energy system operation. For example, renewable energy such as wind is intermittent and stochastic. Modelling energy systems in coarse time resolution cannot reflect the actual requirement of operating reserve resources. This implies that even if sufficient capacity of generators is invested, the energy system may still not be balanced. Alternatively, if we only have a long-term energy balance view, the economic value of fast-ramping generators (such as gas-fired generating units) cannot be adequately reflected, which may result in misleading investment decisions. Therefore, additionally features (such as on-off state and ramping speed) of technologies (such as generators, electrolysers, etc.) need to be embedded in the model.

(3) Practical energy system models should be embedded in the national energy model or at least as a separate model to enable a more detailed analysis. To correctly reflect the value of the hydrogen features which will only be reflected during high-resolution operation horizon, a more practical energy system model, such as unit commitment, economic dispatch, and emergency control, should be embedded or deployed as separate modules to ensure the feasibility of the final decisions during the practical operation. Moreover, with the options in network-related technologies, such as hydrogen blending, a spatial-based energy system model is essential to determine the optimal allocation of resources.

(4) Advanced uncertainty quantification and risk-hedging optimization framework should be developed or employed. The increasing penetration of renewables and evolving energy system structures will cause heterogeneous short-term and long-term uncertainties. These uncertainties comprise fluctuations in renewable energy generation, effects of climate change on renewable resources, variations in demand, policy changes, and advancements in technology. This is particularly critical for national energy system models that are focused on developing long-term roadmaps extending to 2050 or 2060. Consequently, it is important to develop advanced uncertainty-oriented optimization frameworks, which may include scenario-based stochastic optimization, worst-case-based robust optimization, and probability density based distributionally robust optimization.

*5.2 Suggestions for storage*

(1) Enhancing hydrogen storage simulation models

Models should account for diverse hydrogen storage possibilities, including, but not limited to, underground salt cavern storage, depleted oil and gas fields, facilities constructed as needed, and hybrid storage or so-called linepack systems. With unique features for each method from the above-mentioned approaches, this is where models help in providing the much-needed versatility to balance energy demand and supply, especially during peak energy demand over several periods. The simulation models need to be able to represent different types of scenarios of hydrogen storage, considering the metrics including capacity, efficiency, and operational dynamics. For example, risks and technical challenges for underground storage systems should be minimized by taking into account geological features and permeability while determining parameters used in models. Given advantages of offshore storage (e.g., salt caverns and depleted gas fields under the sea), these options should be included in models as feasible pathways to long-term storage. Strict advanced simulations must also be integrated to study structures, as well as environmental effects for such sites.

Adopting these recommendations will enable models to optimize hydrogen energy networks and improve hydrogen energy integration into the growth of renewable resources and integrated energy systems. Following these suggestions can enhance the efficiency, safety, and cost-effectiveness of hydrogen storage systems. These practices are grounded in advanced material science and real-time monitoring systems to ensure optimal performance and long-term reliability in the hydrogen economy.

(2) Incorporate support mechanisms when modeling hydrogen storage

Hydrogen storage involves high investments, and the models could apply mechanisms for governmental support and regional policies to deduct capital costs via subsidies or tax incentives for developing these technologies. Such policies can help enable the hydrogen storage infrastructure buildout across multiple markets.

*5.3 Suggestions for transportation*

As hydrogen is gradually emerging, transportation infrastructure has not been fully established yet and will take time to materialize. Energy systems model should therefore:

(1) Leverage existing infrastructure for near-term solutions: Models should prioritize repurposing existing natural gas pipelines for hydrogen transport in the short term. This approach reduces initial capital requirements, facilitating earlier deployment of hydrogen transport while dedicated infrastructure develops.

(2) Adopt dedicated hydrogen pipelines in the long-term:
- For future scenarios, models should allow for the gradual development of dedicated hydrogen pipelines within and between regions. This approach supports a scalable hydrogen network that aligns with long-term demand growth and regional integration needs.
- In both repurposed and dedicated infrastructure scenarios, capital costs should be carefully integrated to accurately represent the financial investment needed for each phase.

(3) Incorporate hydrogen carriers as a flexible transport option
- Recognizing the challenges in pipeline establishment, models should include hydrogen carriers—such as ammonia, methane, and liquid organic hydrogen carriers—as alternative transport methods. This adds flexibility in meeting diverse demands and infrastructure conditions, especially for regions or applications where pipeline construction is unfeasible.
- Models should account for the costs of carrier production, conversion, transportation, and decomposition at the delivery point, ensuring comprehensive cost analysis across transport options.

*5.4 Suggestions for hydrogen trade*

(1) Strike a balance between domestic hydrogen production and trade

Given the emerging global hydrogen market, models should incorporate both domestic production and international import and export options. Countries with abundant renewable resources may serve as future hydrogen export hubs, offering green hydrogen at competitive prices. Thus, models should account for region-specific hydrogen production costs, transportation logistics, and import costs to capture supply-demand imbalances.

(2) Develop dynamic integration with the national energy models

In current practices, international hydrogen trading models are developed independently from the national energy/hydrogen model. It is understandable that each model addresses distinct scenarios and require varying data inputs. However, with the development of cross-border hydrogen exchanging infrastructures and the ever-increasing role of import/export hydrogen in the energy mix of some countries, the interplay between international and domestic hydrogens has become more noticeable. Therefore, a more dynamic integration of international and national hydrogen models is required, focusing on mutual impacts, such as the supply and demand elasticity and equilibrium of price in the market.

(3) Parameterize hydrogen exports in resource-rich regions

Models in hydrogen-exporting countries or regions (e.g., Australia, the EU, and Ireland) should consider export-specific parameters, such as maximum export volumes, hydrogen type (green or blue), and electricity costs. For instance, Australia's hydrogen export modelling tool demonstrates how models can assess the investment-return potential of hydrogen exports through customizable scenarios.

(4) Develop richer hydrogen market operation models

The current international hydrogen models are formulated based on the match of supply and demand quantity and price in a centralised market. As the hydrogen market matures, transitional support mechanisms may emerge, such as hydrogen contract for difference and subsidy schemes in the UK. This means, that the hydrogen price does not solely depend on the production costs, which will further affect the competitiveness in the international market. Moreover, sophisticated hydrogen trading mechanisms may develop such as bilateral contracts, forward trading, and other secondary markets.

(5) Include infrastructure links and transportation costs

Integrate transportation and storage infrastructure parameters to reflect the complete costs of both imports and exports, from pipeline and shipping costs to storage options. By incorporating these links, models can offer a more comprehensive picture of total hydrogen supply costs and support robust investment and policy planning.

(6) Refine cost calculation methods

In the current international hydrogen trading models, the cost models are coarse. On the production side, climate parameters (such as geo-related wind potential), impacts of human activities (such as shipping density on the power density of wind farms), constraints by onshore energy systems (such as impacts of flexibilities and power system inertia on the hydrogen production power in high time-resolution) should be taken into consideration. On the transmission side, various transportation methods (pipeline/shipping with different hydrogen derivatives), transportation routes and related costs, should be also considered in the model, to offer a more comprehensive picture of total hydrogen supply costs and support robust investment and policy planning.

*5.5 Suggestions for demand*

To accurately capture end-use hydrogen demand, energy system models should incorporate a detailed, sector-specific representation of demand, tailored to the projected demand sources within the targeted jurisdiction. This disaggregation should encompass key sectors such as electricity, industry, residential, transportation, and refining, reflecting each sector's unique demand characteristics and anticipated growth. Additionally, incorporating an hourly or high-resolution temporal representation of end-use demand is essential for capturing fluctuations and peak requirements, thereby enabling models to simulate real-world dynamics more accurately.

*5.6 Suggestions for policies*

Models should include mechanisms to account for region-specific policy supports, such as direct grants and tax credits. Developed economies may favour capital cost reductions through grants, while emerging markets and developing economies (EMDEs) may prioritize tax credits to encourage sector growth. By incorporating these diverse policy mechanisms, models can accurately simulate hydrogen investment attractiveness and policy-driven sector development.

## 6. Conclusion

The aim of this paper is to assess the representation of hydrogen in selected national energy system models, identify modeling gaps, and provide several suggestions for energy analysts and future researchers. In contrast to recent review attempts, this paper focuses on energy systems models that have contributed to some extent to furthering national hydrogen efforts. The paper adopts a comprehensive review approach to identify suitable models to review. The study first identifies countries that have published national hydrogen strategies or have hydrogen plans, with a total of 43 countries identified. Each national hydrogen document was reviewed to identify references to modeling frameworks. The national hydrogen documents were supplemented with energy strategy documents and government publications in cases where the national hydrogen document failed to reference a modeling framework. Models were selected on the basis on whether they were used for hydrogen policy assessments and if accessible documentation was provided. In addition, selected review papers were also reviewed to identify suitable models.

In most of the reviewed models, the general trend is to adopt a simplified representation of the hydrogen infrastructure. This approach allows models to strike a balance between accuracy and preserving computational resources. In particular, the models opt for limited hydrogen production option, gaseous and underground hydrogen storage and pipeline transport. Zhang et al. (2025) arrives at a similar conclusion and suggests that alternative hydrogen technologies are not incorporated into existing energy system models due to low technology readiness or a negligible share in the overall hydrogen sector. From a modeling strategy perspective, most models provide high-level disaggregation of end-use sectors with high spatial and temporal resolutions. Surprisingly, the review finds that most models factor in some variant of hydrogen policies either as a specific policy constraint or a scenario.

The outcome of the paper is a set of suggestions for modeling various aspects of the hydrogen supply chain. It is important to highlight that the review design restricts the assessment to a specific set of models and tools, which constitute the main limitation of this work. Subsequent research could aim to broaden the scope by including additional energy system models that were not addressed in this review. Furthermore, we recognize that customized hydrogen tools either exist (Bush et al., 2013; Penev et al., 2018; Bush et al., 2019) or were in development (Energy Reform et al., 2025) while writing this review that were not adopted in this review. Future researchers might also consider reviewing such models.

**Declaration of competing interests**

The authors have no conflicts of interest to declare.

**Data availability**

 Data will be made available on request.

# References

Aakko-Saksa, P.T., Cook, C., Kiviaho, J., Repo, T., 2018. Liquid organic hydrogen carriers for transportation and storing of renewable energy – Review and discussion. J. Power Sources 396, 803–823. https://doi.org/10.1016/j.jpowsour.2018.04.011

AFRY, 2023. Offshore Renewables Surplus Potential WS3 – Renewable Hydrogen.

Australian Government, 2024. AusH2 - Australia's Hydrogen Opportunities Tool.

Aziz, M., 2021. Liquid Hydrogen: A Review on Liquefaction, Storage, Transportation, and Safety. Energies 14. https://doi.org/10.3390/en14185917

Balyk, O., Andersen, K., Dockweiler, S., Gargiulo, M., Karlsson, K., Næraa, R., Petrović, S., Tattini, J., Termansen, L.B., Venturini, G., 2019. TIMES-DK: Technology-rich multi-sectoral optimisation model of the Danish energy system. Energy Stud. Rev. 23, 13–22. https://doi.org/10.1016/j.esr.2018.11.003

Balyk, O., Glynn, J., Aryanpur, V., Gaur, A., McGuire, J., Smith, A., Yue, X., Daly, H., 2022. TIM: modelling pathways to meet Ireland's long-term energy system challenges with the TIMES-Ireland Model (v1.0). Geosci. Model Dev. 15, 4991–5019. https://doi.org/10.5194/gmd-15-4991-2022

Bistline, J., Cole, W., Damato, G., DeCarolis, J., Frazier, W., Linga, V., Marcy, C., Namovicz, C., Podkaminer, K., Sims, R., Sukunta, M., Young, D., 2020. Energy storage in long-term system models: a review of considerations, best practices, and research needs. Prog. Energy 2. https://doi.org/10.1088/2516-1083/abab68

Blanco, H., Leaver, J., Dodds, P.E., Dickinson, R., García-Gusano, D., Diego, I., Lind, A., Wang, C., Danebergs, J., Baumann, M., 2022. A taxonomy of models for investigating hydrogen energy systems. Renew. Sustain. Energy Rev. 167. https://doi.org/10.1016/j.rser.2022.112698

Bolat, P., Thiel, C., 2014. Hydrogen supply chain architecture for bottom-up energy systems models. Part 1: Developing pathways. Int. J. Hydrog. Energy 39, 8881–8897. http://dx.doi.org/10.1016/j.ijhydene.2014.03.176

Burke, A., Ogden, J., Fulton, L., Cerniauskas, S., 2024. Hydrogen Storage and Transport: Technologies and Costs (No. UCD-ITS-RR-24-17). University of California Davis Institute of Transportation Studies.

Bush, B., Melaina, M., Penev, M., Daniel, W., 2013. SERA Scenarios of Early Market Fuel Cell Electric Vehicle Introductions: Modeling Framework, Regional Markets, and Station Clustering (No. NREL/TP-5400-56588). National Renewable Energy Laboratory (NREL).

Bush, B., Muratori, M., Hunter, C., Zuboy, J., Melaina, M., 2019. Scenario Evaluation and Regionalization Analysis (SERA) Model: Demand Side and Refueling Infrastructure Buildout (No. NREL/TP-5400-70090). National Renewable Energy Laboratory (NREL).

Cabal, H., Lechón, Y., García, D., Gargiulo, M., Labriet, M., Tosato, G., 2012. Description of the update TIMES-Spain model, Project COMET.

CER, 2024. Canada's Energy Future 2023 - Modeling Methods [WWW Document]. Can. Energy Regul. URL https://www.cer-rec.gc.ca/en/data-analysis/canada-energy-future/2023-modeling-methods/ (accessed 7.24.24).

Chu, C., Wu, K., Luo, B., Cao, Q., Zhang, H., 2023. Hydrogen storage by liquid organic hydrogen carriers: Catalyst, renewable carrier, and technology – A review. Carbon Resour. Convers. 6, 334–351. https://doi.org/10.1016/j.crcon.2023.03.007

Collins, S., Deane, J.P., Poncelet, K., Panos, E., Pietzcker, R.C., Ó Gallachóir, B.P., 2017. Integrating short term variations of the power system into integrated energy system models: A methodological review. Renew. Sustain. Energy Rev. 76, 839–856. http://dx.doi.org/10.1016/j.rser.2017.03.090

Cosmi, C., Leo, S.D., Loperte, S., Macchiato, M., Pietrapertosa, F., Salvia, M., Cuomo, V., 2009. A model for representing the Italian energy system: The NEEDS-TIMES experience. Renew. Sustain. Energy Rev. 13, 763–776.

Czyżak, P., Mańko, M., Sikorski, M., Stępień, K., Wieczorek, B., 2021. An open energy model of the Polish power sector based on the PyPSA framework. Instrat, Warsaw, Poland.

Dagdougui, H., 2012. Models, methods and approaches for the planning and design of the future hydrogen supply chain. Int. J. Hydrog. Energy 37, 5318–5327. https://doi.org/doi:10.1016/j.ijhydene.2011.08.041

Daghia, F., Baranger, E., Tran, D.T., Pichon, P., 2020. A hierarchy of models for the design of composite pressure vessels. Compos. Struct. 235. https://doi.org/10.1016/j.compstruct.2019.111809

Daly, H.E., Fais, B., 2014. UK TIMES Model Overview. UCL Energy Institute.

Danebergs, J., Rosenberg, E., Seljom, P.M.S., Kvalbein, L., Haaskjold, K., 2022. Documentation of IFE-TIMES-Norway v2 (No. IFE/E-2021/005). Institute for Energy Technology.

DeCarolis, J., Daly, H., Dodds, P., Keppo, I., Li, F., McDowall, W., Pye, S., Strachan, N., Trutnevyte, E., Usher, W., Winning, M., Yeh, S., Zeyringer, M., 2017. Formalizing best practice for energy system optimization modelling. Appl. Energy 194, 184–198. http://dx.doi.org/10.1016/j.apenergy.2017.03.001

DECHEMA, acatech, 2024. Comparative analysis of international hydrogen strategies, Country analysis 2023. Frankfurt, Germany.

DESNZ, 2023. Hydrogen Production Business Model /Net Zero Hydrogen Fund: HAR1 successful projects [WWW Document]. Dep. Energy Secur. Net Zero. URL https://www.gov.uk/government/publications/hydrogen-production-business-model-net-zero-hydrogen-fund-shortlisted-projects/hydrogen-production-business-model-net-zero-hydrogen-fund-har1-successful-projects (accessed 9.11.24).

DESNZ, 2021. Hydrogen Analytical Annex. Department for Energy Security and Net Zero (DESNZ).

DESNZ, BEIS, 2021. Industrial decarbonisation strategy (No. CP 399). Department for Energy Security and Net Zero (DESNZ); Department for Business, Energy & Industrial Strategy (BEIS).

Dodds, P.E., 2021. Review of the Scottish TIMES energy system model. ClimateXChange, Edinburgh, United Kingdom.

E3M, 2018. PRIMES Model Version 2018: Detailed model description. E3-Modelling.

Ea Energianalyse, 2018. Balmorel User Guide.

EIA, 2024. Requirements for the Hydrogen Market Module in the National Energy Modeling System. U.S. Energy Information Administration.

Ekhtiari, A., Flynn, D., Syron, E., 2022. Green Hydrogen Blends with Natural Gas and Its Impact on the Gas Network. Hydrogen 3, 402–417. https://doi.org/10.3390/hydrogen3040025

Ekhtiari, A., Flynn, D., Syron, E., 2020. Investigation of the Multi-Point injection of green hydrogen from curtailed renewable power into a gas network. Energies 13. https://doi.org/10.3390/en13226047

Element Energy, 2021. Net-Zero Industrial Pathways (N-ZIP) Model. BEIS Climate Change Committee.

Element Energy, 2020. Deep-Decarbonisation Pathways for UK Industry. BEIS Climate Change Committee, Cambridge, United Kingdom.

Elgowainy, A., Reddi, K., Mintz, M., Brown, D., 2015. H2A Delivery Scenario Analysis Model Version 3.0 (HDSAM 3.0) User's Manual. U.S. Department of Energy (DOE).

Energy Reform, MullanGrid, Wind Energy Ireland, 2025. SPINE H2-IRL 2023-202 Derivable 5.3 Final Report. Energy Reform, Dublin, Ireland.

Fajardy, M., Morris, J., Gurgel, A., Herzog, H., Mac Dowell, N., Paltsev, S., 2021. The economics of bioenergy with carbon capture and storage (BECCS) deployment in a 1.5 °C or 2 °C world. Glob. Environ. Change 68, 102262. https://doi.org/10.1016/j.gloenvcha.2021.102262

Faye, O., Szpunar, J., Eduok, U., 2022. A critical review on the current technologies for the generation, storage, and transportation of hydrogen. Int. J. Hydrog. Energy 47, 13771–13802. https://doi.org/10.1016/j.ijhydene.2022.02.112

Frieden, F., Leker, J., 2024. Future costs of hydrogen: a quantitative review. Sustain. Energy Fuels 8, 1806–1822. https://doi.org/10.1039/D4SE00137K

Gaeta, M., 2013. Times-Italia: elaborazione e analisi di Scenari per il Sistema Energetico Nazionale. ENEA-Unità Centrale Studi e Strategie, Bologna, Italy.

Gayathri, R., Mahboob, S., Govindarajan, M., Al-Ghanim, K.A., Ahmed, Z., Al-Mulhm, N., Vodovnik, M., Vijayalakshmi, S., 2021. A review on biological carbon sequestration: A sustainable solution for a cleaner air environment, less pollution and lower health risks. J. King Saud Univ. - Sci. 33. https://doi.org/10.1016/j.jksus.2020.101282

Gouveia, J.P., Dias, L., Seixas, J., 2012. TIMES_PT: Integrated Energy System Modeling, in: Proceedings of the First International Workshop on Information Technology for Energy Applications. Presented at the Information Technology for Energy Applications, Lisbon, Portugal, pp. 69–78.

Government of South Australia, 2020. South Australia: A global force in Hydrogen.

Hanley, E.S., Deane, J., Gallachóir, B.Ó., 2018. The role of hydrogen in low carbon energy futures–A review of existing perspectives. Renew. Sustain. Energy Rev. 82, 3027–3045. https://doi.org/10.1016/j.rser.2017.10.034

Hosseini, S.E., 2023. Hydrogen storage and delivery, in: Fundamentals of Hydrogen Production and Utilization in Fuel Cell Systems. Elsevier.

Hughes, A., Merven, B., McCall, B., Ahjum, F., Caetano, T., Hartley, F., Ireland, G., Burton, J., Marquard, A., 2021. Evolution, Assumptions and Architecture of the South African Energy Systems Model SATIM. Energy Systems Research Group Working Paper.

Hydrogen Council, McKinsey & Company, 2023. Hydrogen Insights 2023: The state of the global hydrogen economy, with a deep dive into renewable hydrogen cost evolution.

Hydrogen UK, 2024. Hydrogen Import and Export: Unlocking the UK's hydrogen trade potential.

Hydrogen UK, 2023. Recommendations for the Acceleration of Hydrogen Networks. Hydrogen UK.

Hydrogen UK, ENA, 2023. International Hydrogen Progress Index: Hydrogen Networks. Hydrogen UK; Energy Networks Association.

IEA, 2024. Global Hydrogen Review 2024. International Energy Agency (IEA), Paris, France.

IEA, 2023a. Global Hydrogen Review 2023. International Energy Agency (IEA), Paris, France.

IEA, 2023b. Global Energy and Climate Model. International Energy Agency (IEA), Paris, France.

IEA, 2021a. Net Zero by 2050: A Roadmap for the Global Energy Sector. International Energy Agency (IEA).

IEA, 2021b. Ammonia technology roadmap: Towards more sustainable nitrogen fertiliser production. International Energy Agency (IEA), Paris, France.

IEA, 2019. The Future of Hydrogen: Seizing today's opportunities, Report prepared by the IEA for the G20, Japan. International Energy Agency (IEA).

IRENA, 2024a. Green hydrogen strategy: A guide to design. International Renewable Energy Agency (IRENA), Masdar City, UAE.

IRENA, 2024b. International Co-operation to accelerate green hydrogen deployment. International Renewable Energy Agency (IRENA), Abu Dhabi, UAE.

IRENA, 2023. World Energy Transitions Outlook 2023: 1.5°C Pathway. International Renewable Energy Agency (IRENA), Masdar City, UAE.

IRENA, 2021. Green hydrogen supply: A guide to policy making. International Renewable Energy Agency (IRENA), Abu Dhabi, UAE.

Kannan, R., Turton, H., 2014. Switzerland Energy Transition Scenarios – Development and Application of the Swiss TIMES Energy System Model (STEM) (No. 14– 06). Paul Scherrer Institut (PSI).

Kountouris, I., Bramstoft, R., Madsen, T., Gea-Bermúdez, J., Münster, M., Keles, D., 2024. A unified European hydrogen infrastructure planning to support the rapid scale-up of hydrogen production. Nat. Commun. 15.

Laguna, J.C., Moglianesi, A., Vingerhoets, P., Lodewijks, P., 2022. Description of the EnergyVille TIMES Be model. EnergyVille.

Langer, L., Weibezahn, J., Giehl, J.F., Neumann, F., Göke, L., Kountouris, I., Münster, M., Thorendahl, A.V., Hartvig, M., Eleftheriou, D., Bramstoft, R., 2024. Hydrogen infrastructure modeling in macro-energy systems models - Lessons learned, best practices, and potential next steps (workshop insights). Int. J. Hydrog. Energy 70, 629–634. https://doi.org/10.1016/j.ijhydene.2024.05.137

Li, L., Manier, H., Manier, M.-A., 2019. Hydrogen supply chain network design: An optimization-oriented review. Renew. Sustain. Energy Rev. 103, 342–360. https://doi.org/10.1016/j.rser.2018.12.060

Lugovoy, O., 2007. Towards a new Russia TIMES Model. Environmental Defense.

Lund, H., Thellufsen, J.Z., 2024. EnergyPLAN Advanced Energy Systems Analysis Computer Model Documentation Version 16.3. Aalborg University.

Maryam, S., 2017. Review of modelling approaches used in the HSC context for the UK. Int. J. Hydrog. Energy 42, 24927–24938. https://doi.org/10.1016/j.ijhydene.2017.04.303

Matteo, N., 2020. A TIMES-like open-source model for the Italian energy system.

McKinsey & Company, 2023. Global Hydrogen Flows - 2023 Update.

McKinsey & Company, 2022. Global Hydrogen Flows: Hydrogen trade as a key enabler for efficient decarbonization.

Michael, M., Benjamin, M., Aaron, S., 2017. NREL Offshore Balance-of-System Model.

Muhammed, N.S., Haq, B., Al Shehri, D., Al-Ahmed, A., Rahman, M.M., Zaman, E., 2022. A review on underground hydrogen storage: Insight into geological sites, influencing factors and future outlook. Energy Rep. 8, 461–499. https://doi.org/10.1016/j.egyr.2021.12.002

Mulky, L., Srivastava, S., Lakshmi, T., Sandadi, E.R., Grour, S., Thomas, N.A., Priya, S.S., Sudhakar, K., 2024. An overview of hydrogen storage technologies – Key challenges and opportunities. Mater. Chem. Phys. 325. https://doi.org/10.1016/j.matchemphys.2024.129710

Netherlands 2030, 2023. Klimaat- en Energieverkenning 2022.

Neumann, F., Zeyen, E., Victoria, M., Brown, T., 2023. The potential role of a hydrogen network in Europe. Joule 7, 1793–1817. https://doi.org/10.1016/j.joule.2023.06.016

Nicoli, M., 2024. TEMOA-Italy.

Osman, A.I., Nasr, M., Eltaweil, A.S., Hosny, M., Farghali, M., Al-Fatesh, A.S., Rooney, D.W., Abd El-Monaem, E.M., 2024. Advances in hydrogen storage materials: harnessing innovative technology, from machine learning to computational chemistry, for energy storage solutions. Int. J. Hydrog. Energy 67, 1270–1294. https://doi.org/10.1016/j.ijhydene.2024.03.223

Oxford Institute for Energy Studies, 2023. Renewable Hydrogen Import Routes into the EU.

Papadias, D.D., Peng, J.-K., Ahluwalia, R.K., 2021. Hydrogen carriers: Production, transmission, decomposition, and storage. Int. J. Hydrog. Energy 46, 24169–24189. https://doi.org/10.1016/j.ijhydene.2021.05.002

Patonia, A., Poudineh, R., 2022. Global trade of hydrogen: what is the best way to transfer hydrogen over long distances? Oxford Institute for Energy Studies Paper.

Penev, M., Saur, G., Hunter, C., Zuboy, J., 2018. H2A Hydrogen Production Model: Version 3.2018 User Guide. National Renewable Energy Laboratory (NREL).

Raj, A., Larsson, I.A.S., Ljung, A.-L., Forslund, T., Andersson, R., Sundström, J., Lundström, T.S., 2024. Evaluating hydrogen gas transport in pipelines: Current state of numerical and experimental methodologies. Int. J. Hydrog. Energy 67, 136–149. https://doi.org/10.1016/j.ijhydene.2024.04.140

Ravn, H., 2011. Balmorel: Getting started.

Reddi, K., Elgowainy, A., Rustagi, N., Gupta, E., 2018. Techno-economic analysis of conventional and advanced high-pressure tube trailer configurations for compressed hydrogen gas transportation and refueling. Int. J. Hydrog. Energy 43, 4428–4438. https://doi.org/10.1016/j.ijhydene.2018.01.049

Richard, L., 2016. Documentation for the TIMES Model PART I.

Riekkola, A.K., 2015. National Energy System Modelling for Supporting Energy and Climate Policy Decision-making: The Case of Sweden (PhD Thesis). Chalmers University of Technology, Göteborg, Sweden.

Riera, J.A., Lima, R.M., Knio, O.M., 2023. A review of hydrogen production and supply chain modeling and optimization. Int. J. Hydrog. Energy 48, 13731–13755. https://doi.org/10.1016/j.ijhydene.2022.12.242

Riester, C.M., García, G., Alayo, N., Tarancón, A., Santos, D.M.F., Torrell, M., 2022. Business Model Development for a High-Temperature (Co-)Electrolyser System. Fuels 3, 392–407. https://doi.org/10.3390/fuels3030025

SARI/EI, 2016. India TIMES Model (Electricity)- Key Inputs & Results. South Asia Regional Initiative for Energy Integration (SARI/EI), New Delhi, India.

Scottish Government, 2018. Climate Change Plan: The Third Report on Proposals and Policies 2018-2032 Technical Annex (No. SG/2018/18). The Scottish Government.

Song, S., Lin, H., Sherman, P., Yang, X., Nielsen, C.P., Chen, X., McElroy, M.B., 2021. Production of hydrogen from offshore wind in China and cost-competitive supply to Japan. Nat. Commun. 12, 6953. https://doi.org/10.1038/s41467-021-27214-7

Statista Research Department, 2024. Forecast production share of hydrogen worldwide in 2050, by technology [WWW Document]. URL https://www.statista.com/statistics/1364669/forecast-global-hydrogen-production-share-by-technology/

VITO, 2013. The Belgian TIMES model. Vlaamse Instelling voor Technologisch Onderzoek (VITO).

Wang, C., Zhou, D., Xiao, W., Shui, C., Ma, T., Chen, P., Hao, J., Yan, J., 2023. Research on the dynamic characteristics of natural gas pipeline network with hydrogen injection considering line-pack influence. Int. J. Hydrog. Energy 48, 25469–25486. https://doi.org/10.1016/j.ijhydene.2023.03.298

Zhang, T., Qadrdan, M., Wu, J., Couraud, B., Stringer, M., Walker, S., Hawkes, A., Allahham, A., Flynn, D., Pudjianto, D., Dodds, P., Strbac, G., 2025. A systematic review of modelling methods for studying the integration of hydrogen into energy systems. Renew. Sustain. Energy Rev. 208. https://doi.org/10.1016/j.rser.2024.114964

Züttel, A., 2004. Hydrogen storage methods. Naturwissenschaften 91, 157–172. https://doi.org/10.1007/s00114-004-0516-x

Züttel, A., 2003. Materials for hydrogen storage. Mater. Today 6, 24–33. https://doi.org/10.1016/S1369-7021(03)00922-2